%% file: main.tex
\documentclass[11pt,a4paper]{article}
\usepackage[top=2.2cm,bottom=2.4cm,left=2.4cm,right=2.4cm,footskip=26pt]{geometry}
\usepackage[T1]{fontenc}
\usepackage{lmodern,microtype}
\usepackage{helvet}
\usepackage[hyphens]{url}
\usepackage{graphicx}
\usepackage[numbers,sort&compress]{natbib}
\usepackage{caption,titlesec}
\usepackage{amsmath,amssymb,xspace,booktabs,multirow,tabularx}
\usepackage[table]{xcolor}
\usepackage{newfloat,placeins,float,wrapfig,needspace}
\newenvironment{evidencecolumn}[1]{\begin{minipage}[t]{#1\linewidth}\vspace{0pt}}{\end{minipage}}
\newsavebox{\rqfourtable}
\newcommand{\sidecaption}[2]{\captionsetup{font=footnotesize,skip=4pt,justification=raggedright}\captionof{#1}{#2}}
\usepackage[most]{tcolorbox}
\usepackage{hyperref}
\definecolor{reportInk}{HTML}{0F172A}
\definecolor{reportMuted}{HTML}{475569}
\definecolor{reportLine}{HTML}{CBD5E1}
\definecolor{reportPaper}{HTML}{F1F5F9}
\definecolor{reportBlue}{HTML}{2563EB}
\hypersetup{colorlinks,allcolors=reportBlue,pdftitle={Change2Task: From Repository Changes to Executable Coding Agent Tasks and Environments}}
\titleformat*{\section}{\Large\sffamily\bfseries\color{reportInk}}
\titleformat*{\subsection}{\large\sffamily\bfseries\color{reportInk}}
\titleformat*{\subsubsection}{\normalsize\sffamily\bfseries\color{reportInk}}
\titlespacing*{\section}{0pt}{1.2em}{0.6em}
\titlespacing*{\subsection}{0pt}{0.9em}{0.4em}
\DeclareFloatingEnvironment[fileext=loa,name=Algorithm,placement=tb]{algorithm}

\definecolor{c2tNavy}{HTML}{315F7D}
\definecolor{c2tBlueBg}{HTML}{EDF5F9}
\definecolor{c2tTeal}{HTML}{3A8586}
\definecolor{c2tTealBg}{HTML}{EDF8F6}
\definecolor{c2tOrange}{HTML}{C97542}
\definecolor{c2tOrangeBg}{HTML}{FFF4EB}
\definecolor{c2tPurple}{HTML}{755593}
\definecolor{c2tPurpleBg}{HTML}{F5F0F8}
\newtcolorbox{promptbox}[3]{enhanced,breakable,colback=reportPaper,colframe=reportLine,
  colbacktitle=reportMuted,coltitle=white,fonttitle=\bfseries,fontupper=\small,
  title=#1,arc=2.5mm,outer arc=2.5mm,boxrule=0.55pt,
  left=1.8mm,right=1.8mm,top=1.2mm,bottom=1.2mm,
  before skip=3pt,after skip=3pt,segmentation style={solid,#2}}
\newtcolorbox{casebox}[1]{enhanced,breakable,colback=reportPaper,
  colframe=reportLine,colbacktitle=reportMuted,coltitle=white,
  fonttitle=\bfseries,fontupper=\small,title=#1,arc=2.5mm,
  outer arc=2.5mm,boxrule=0.55pt,left=1.8mm,right=1.8mm,
  top=1.2mm,bottom=1.2mm,before skip=3pt,after skip=3pt}
\newtcolorbox{findingbox}[1]{enhanced,colback=reportPaper,colframe=reportMuted,
colbacktitle=reportMuted,coltitle=white,title={#1},fonttitle=\bfseries,
boxrule=.7pt,arc=3mm,left=3mm,right=3mm,top=2mm,bottom=2mm,
before={\par\nopagebreak[4]\vspace{7pt}\noindent},after skip=7pt}
\newcommand{\promptfield}[1]{{\itshape #1}}
\newcommand{\sys}{\mbox{\textsc{Change2Task}}\xspace}
\newcommand{\updelta}[1]{\(\uparrow\)#1}
\newcommand{\downdelta}[1]{\(\downarrow\)#1}
\newcommand{\matchedAttempts}{1,130}
\newcommand{\fullRecovered}{900}
\newcommand{\fullRecoveryRate}{79.6\%}
\newcommand{\bugBaselineAttempts}{621}
\newcommand{\oneShotRecovered}{387}
\newcommand{\oneShotRecoveryRate}{62.3\%}

\newcommand{\exactRecoveryRate}{13.0\%}
\newcommand{\recoveryGain}{29.2\%}

\newcommand{\rev}[1]{#1}
\newcommand{\environmentTimeSaving}{58.4\%}
\newcommand{\environmentStorageSaving}{71.2\%}
\newcommand{\constructionCostSaving}{10.8\%}
\newcommand{\historicalEnvironmentTime}{3.60 h}
\newcommand{\sharedEnvironmentTime}{1.50 h}
\newcommand{\historicalEnvironmentStorage}{6.20 GB}
\newcommand{\sharedEnvironmentStorage}{1.79 GB}

\begin{document}

\noindent\includegraphics[height=.75cm]{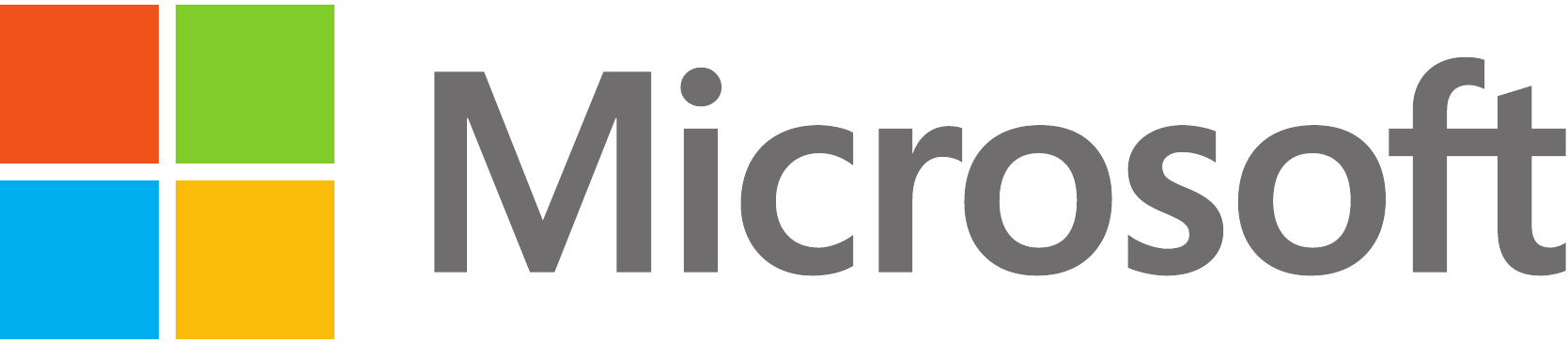}\par
\vspace{3pt}\noindent{\color{reportLine}\rule{\linewidth}{0.8pt}}\par
\begin{center}
{\sffamily\bfseries\fontsize{21}{25}\selectfont\color{reportInk}
Change2Task: From Repository Changes\\
to Executable Coding Agent Tasks\\
and Environments\par}
\vspace{1em}
{\color{reportInk}Haomin Qi\textsuperscript{\textdagger}, Xingliang Wang, Xuanqi Gao, Baihui Sang, Xin Zhang, Minghua Ma,\\
Pengfei Gao, Yu Kang\textsuperscript{\textdagger}, Qingwei Lin, Saravan Rajmohan, Dongmei Zhang, Qi Zhang\par}
\smallskip{\small\color{reportMuted}Microsoft Research\par}
\end{center}
\begin{tcolorbox}[enhanced,frame hidden,colback=reportPaper,arc=8pt,boxrule=0pt,
  left=5mm,right=5mm,top=4mm,bottom=4mm,before skip=4pt,after skip=10pt]
\small\color{reportInk}\raggedright

Coding agents learn from and are measured on executable tasks, but building such tasks from repository history usually means rebuilding an environment at an old revision for every task, which is costly and brittle. We observe that a merged pull request (PR) can instead be undone on today's code: removing the effect of a bug fix or feature from a maintained revision recreates the same task inside an environment that already builds and runs. We present \textbf{Change2Task}, which converts merged PRs into verified tasks on an up-to-date revision of the same repository. It reconstructs the pre-change state by patch reversal, code mapping, or agent-based rewriting, and keeps a task only if its target tests pass, fail, and pass again across the healthy, task, and restored states. From 1{,}130 PRs spanning bug fixing, feature addition, test generation, API migration, and security repair, \textsc{Change2Task} builds 900 verified tasks, recovering 29\% more than SWE-smith's PR Mirror on matched bug-fix candidates. The reconstructed tasks preserve the evaluation signal: in aggregate, four coding agents solve 41.1\% of tasks on both the original and reconstructed versions, with 89.7\% per-task agreement. Sharing modern environments across tasks further reduces setup time by 58\% and storage by 71\%. On 11 private industrial repositories, \textsc{Change2Task} yields 1{,}000 tasks on which the strongest model reaches only 14.6\% Pass@3.
\par\smallskip
{\sffamily\bfseries\color{reportBlue}Code \& Database:}\enspace\href{https://github.com/microsoft/RepoLaunch/tree/change2task}{github.com/microsoft/RepoLaunch/tree/change2task}
\par\smallskip
{\footnotesize\color{reportMuted}\textsuperscript{\textdagger}\,Correspondence to Haomin Qi (\href{mailto:h5qi@ucsd.edu}{h5qi@ucsd.edu}) and Yu Kang (\href{mailto:kangyu159@hotmail.com}{kangyu159@hotmail.com}).}
\end{tcolorbox}

\input{body/Introduction.tex}
\input{body/Background.tex}
\input{body/Methodology.tex}
\input{body/Experiment.tex}
\input{body/Conclusion.tex}

\renewcommand{\bibfont}{\small}
\setlength{\bibsep}{3pt plus 1pt minus 1pt}
\bibliographystyle{unsrtnat}
\bibliography{Reference}

\FloatBarrier
\appendix
\section*{Appendix}
\input{body/Appendix.tex}

\end{document}

%% file: body/Introduction.tex
\section{Introduction}\label{sec:intro}

Coding agents have shifted software automation from single turn generation to iterative interaction with development environments. Systems such as SWE-agent and OpenHands search repositories, edit files, invoke tools, run tests, and revise solutions from execution feedback~\cite{yang2024sweagent,wang2024openhands}. Each executable task therefore acts as a unit of agent data, coupling a repository state with dependencies, tools, a specification, and a verifier. We use \emph{environment engineering} for constructing this executable substrate~\cite{li2026agenticenvironment}. Its supply bounds the scale and diversity of agent training, benchmarks, and continuous evaluation.

Recent systems make this dependence explicit. Kimi-Dev adapts agents from public trajectories, Qwen3-Coder-Next synthesizes verifiable coding tasks with executable environments, and GLM-5 develops asynchronous reinforcement learning infrastructure for long horizon interaction~\cite{yang2025kimidev,cao2026qwen3codernext,glm5team2026}. Producing these environments is expensive: daVinci-Env reports 45,320 Docker environments across more than 12.8K repositories and approximately \$891K in construction cost, while SWE-Universe produces 807,693 verifiable environments using specialized distributed infrastructure~\cite{fu2026openswe,chen2026sweuniverse}. Snapshots dedicated to individual tasks further multiply images, storage, registry traffic, cold starts, and rollout pressure. The central systems question is how to obtain more verified agent data from every executable repository environment.

\rev{Historical benchmarks preserve developer issues, patches, and tests at their source revisions~\cite{jimenez2023swe}. Fresh synthesis scales repair-task generation in prepared repositories, as SWE-smith demonstrates~\cite{yang2025swesmith}. Reconstructing a repository's own historical changes on maintained code retains developer grounding and increases the data obtained from each environment.}

\begin{wrapfigure}{r}{.49\linewidth}
\captionsetup{font=footnotesize,skip=4pt,justification=raggedright}
\centering
    \includegraphics[width=\linewidth]{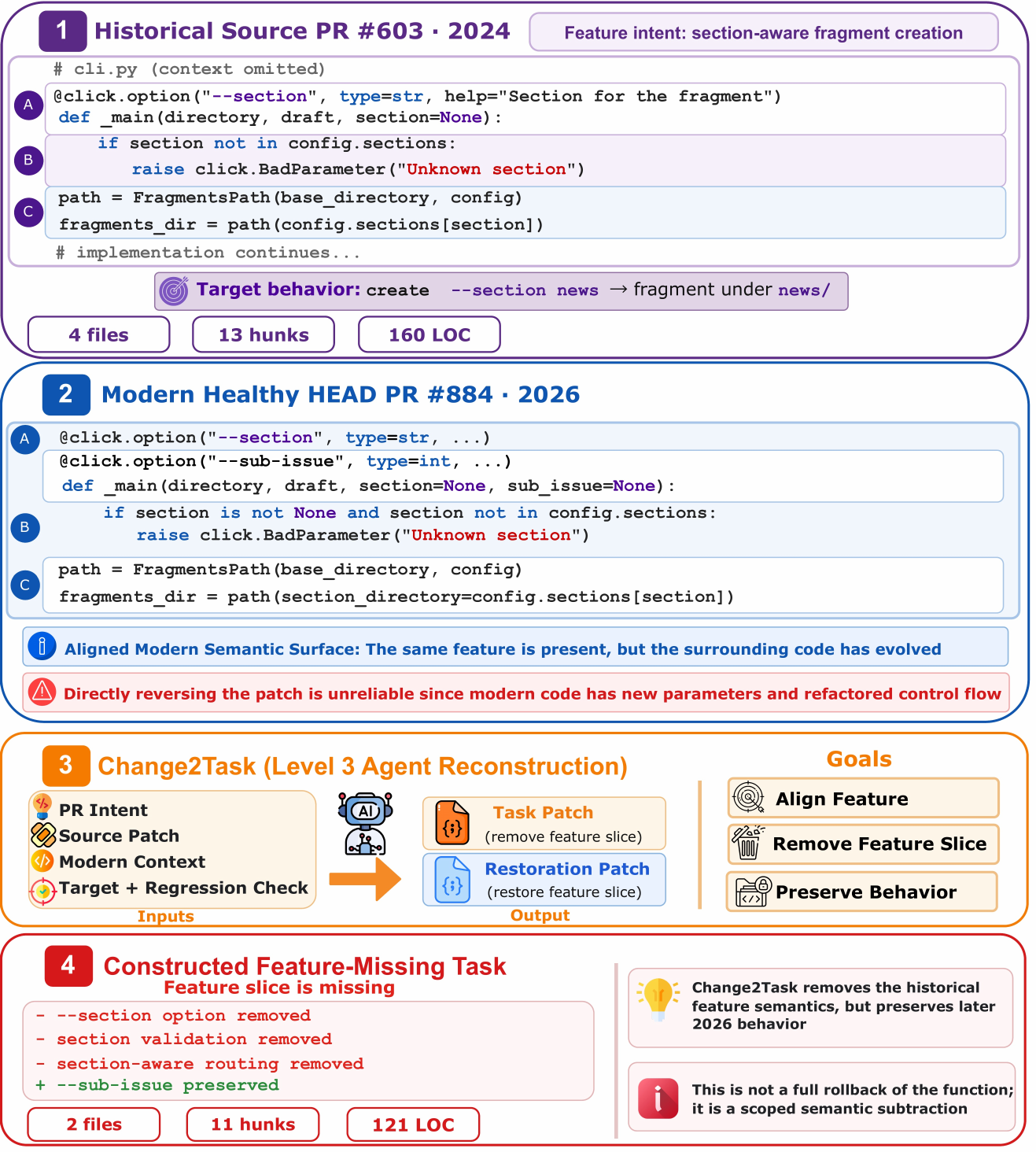}
    \caption{\rev{A Change2Task Feature Addition Task Case Construction example.}}
    \label{fig:concrete-instance}
\end{wrapfigure}

To address this problem, we introduce \sys, a framework and tool for turning repository history into executable coding agent data. It starts from a merged pull request (PR) and a fixed, runnable descendant revision in the same repository. The PR supplies the description, implementation patch, and behavioral evidence; the descendant supplies the healthy modern base. Construction proceeds through Level 1, Patch Reversal; Level 2, Code Mapping; and Level 3, Agent Reconstruction, escalating as repository evolution weakens direct correspondence. Each accepted task patch has a restoration patch and is validated across healthy, task, and restored states. By deriving multiple lightweight task variants from one maintained base, \sys reduces repeated environment setup, storage pressure, and manual construction and validation effort while expanding executable data for coding agent training and task generation.

Task objectives are expressed through adapters that define the goal, expected output, executable oracle, and edit scope. \rev{We evaluate Bug Fix, Feature Addition, Test Generation, Application Programming Interface (API) Migration, and Security Repair; each adapter defines a distinct output and verification contract.} The same interface naturally supports additional coding tasks whenever their maintenance intent and executable success condition can be derived from historical PR evidence.

Figure~\ref{fig:concrete-instance} gives a Feature Addition example. Repository history supplies a task grounded in developer evidence, while a maintained revision supplies the environment in which that task is executed.

Our contribution is \sys, a general framework and implemented tool that transfers developer-grounded maintenance changes to current code and verifies their provenance, behavior, scope, and restoration. From 1,130 eligible changes, \sys finalizes 900 paired tasks; on 621 matched Bug Fix candidates, it reconstructs 500 versus 387 for SWE-smith PR Mirror. \rev{The corpus achieves 0.894 task-weighted source change profile fidelity.} \rev{Under matched agent evaluation, historical and reconstructed cases attain 89.7\% aggregate outcome agreement with consistent agent ranking across the two branches.} Reusing 388 modern bases reduces environment time by \environmentTimeSaving{}, storage by \environmentStorageSaving{}, and end-to-end expenditure by \constructionCostSaving{}.

\rev{We further examine the contributions associated with each research question through matched studies of construction stages, semantic review, branch consistency, and resource use. An industrial deployment constructs 1,000 additional tasks from private PR histories and evaluates four models on a fixed 500-task cohort.}

%% file: body/Background.tex
\Needspace{7\baselineskip}
\section{Related Work}\label{sec:problem}

\subsection{Coding Agents and Repository Environments}

Agents operating at repository scale inspect code, edit files, invoke tools, and revise solutions from execution feedback. SWE-agent studies interfaces between agents and computers, OpenHands provides a sandboxed platform, AutoCodeRover combines program structure with search, and Agentless uses fixed localization, repair, and validation stages~\cite{yang2024sweagent,wang2024openhands,zhang2024autocoderover,xia2025agentless}. SWE-bench established issue resolution on repository snapshots with hidden tests; SWE-Gym and R2E-Gym extend executable feedback to training~\cite{jimenez2023swe,pan2025swegym,jain2025r2egym}. TopoEdge applies topology retrieval and executable protocol checks to network configuration generation and repair~\cite{qi2026topoedge}. Together, these systems show that performance depends on the model, repository state, interaction interface, and exposed verifier.

Recent benchmarks expand freshness, industrial difficulty, and task breadth through Multi-SWE-bench, SWE-bench-Live, SWE-Bench Pro, and SWE-bench Multimodal~\cite{zan2025multi,zhang2025swe,deng2025swe,yang2025swebenchmultimodal}. FeatureBench and SWE-Cycle cover feature development and fuller issue resolution cycles; SWE-EVO and TestEvo-Bench study repository evolution and joint code and test evolution~\cite{zhou2026featurebench,guan2026swecycle,thai2025sweevo,wang2026testevo}. Live, multimodal, and evolutionary settings also require environments to remain executable as code, dependencies, and interfaces change. This breadth increases demand for maintained environments, outputs specific to each task, and reliable verifiers.

\begin{figure}[htbp]
    \centering
    \includegraphics[width=0.96\textwidth]{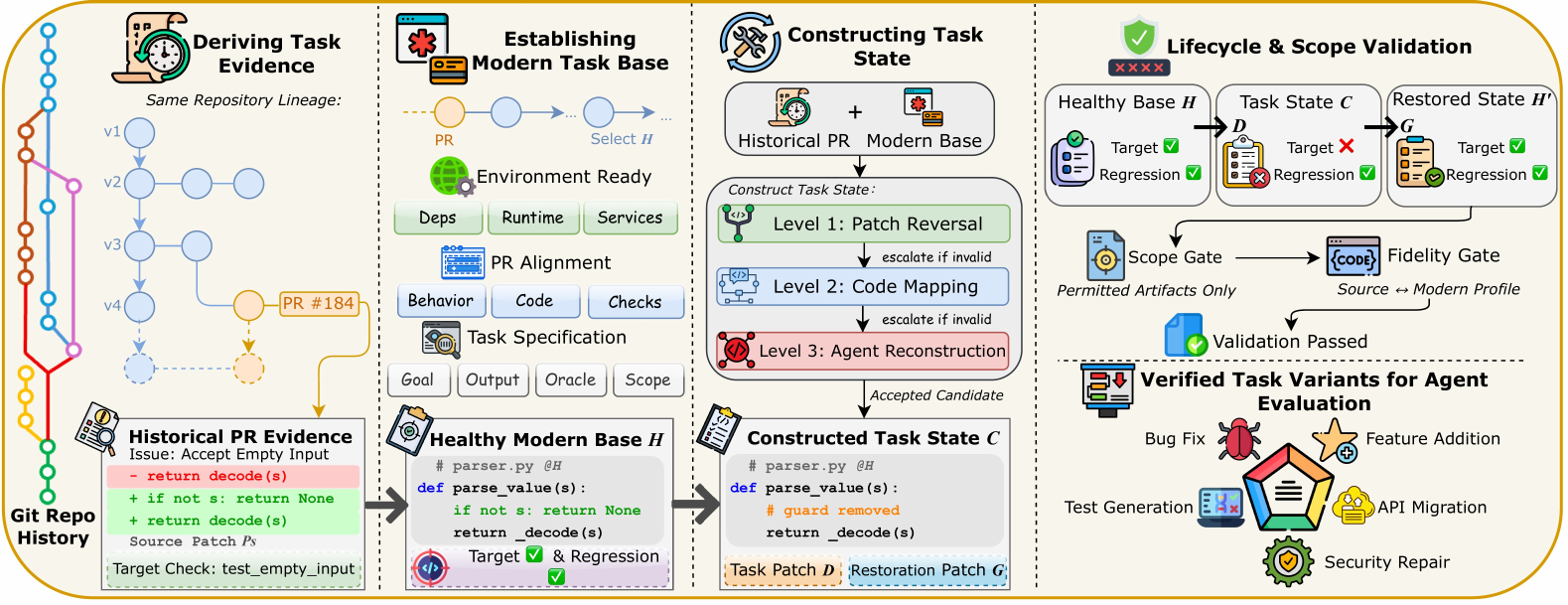}
    \caption{Overview of the \sys workflow.}
    \label{fig:overview}
\end{figure}

\subsection{Executable Repository Infrastructure}

Executable evaluation requires dependencies, services, build commands, test discovery, and isolation. EnvBench and SetupBench document persistent setup obstacles~\cite{eliseeva2025envbench,arora2025setupbench}. Repo2Run synthesizes Dockerfiles from feedback, RepoLaunch automates setup and test management, and MEnvAgent combines planning, verification, and cross-language reuse~\cite{hu2025repo2run,li2026repolaunch,guo2026menvagent}. Their focus is recovering or maintaining a runnable revision, an essential prerequisite for agent evaluation. \rev{Our study measures how many historically grounded task states a prepared base can support.}

At scale, SWE-Factory coordinates environment, test, and instance construction, while daVinci-Env and SWE-Universe combine specialized builders, self-verification, and distributed execution~\cite{guo2025swefactory,fu2026openswe,chen2026sweuniverse}. SWE-smith reuses configured images for synthetic instances~\cite{yang2025swesmith}. These efforts reduce setup cost or increase corpus scale. \sys is complementary: it begins after a healthy base is executable and increases the verified, developer-grounded task supply obtained from that investment.

\subsection{Task Construction for Agents}

Historical benchmarks ground tasks in developer issues, patches, and tests. Defects4J, SWE-bench, and SWE-Gym preserve realistic outcomes at original revisions~\cite{just2014defects4j,jimenez2023swe,pan2025swegym}; SWT-Bench emphasizes tests that distinguish faulty and repaired states~\cite{mundler2024swtbench}. WitnessGym constructs bug-validation tasks through bug injection and retains cases supported by executable witnesses~\cite{qi2026witnessgymbenchmarkingcodingagents}. R2E-Gym and SWE-Synth scale through mutation, model generation, and executable feedback~\cite{jain2025r2egym,pham2025swesynth}. For downstream training, governance-aware multilingual fine-tuning couples model adaptation with data filtering~\cite{qi2025governance}. \rev{Our study grounds task intent in maintenance events recorded by the repository.}

The closest method, SWE-smith's PR Mirror, asks a model to undo a historical PR in a current repository and retains candidates that break target tests~\cite{yang2025swesmith}. FeatureBench removes tested feature units to construct feature-development tasks~\cite{zhou2026featurebench}. \rev{These methods demonstrate task generation inside prepared repositories. Our setting spans three construction routes, repository evolution, and five task objectives with distinct oracles.}

\sys reconstructs a real maintenance event on a maintained descendant of the same repository. Shared lineage and target, regression, restoration, scope, and source change profile checks connect the modern task to its source PR. One validation core then serves multiple objectives without rebuilding the base environment.

%% file: body/Methodology.tex
\section{Methodology}\label{sec:design}

\sys reconstructs, in modern code, the task condition represented by a historical PR. It creates an executable task state on a fixed healthy descendant of the same repository while retaining the PR as developer evidence. The modern revision supplies the environment in which an agent will act.

We use one compact notation to describe this lifecycle:
\[
V_{\mathrm{pre}} \xrightarrow{P_s} V_{\mathrm{post}},
\qquad
H \xrightarrow{D} C \xrightarrow{G} H'.
\]
The source patch $P_s$ changes the historical state before the PR, $V_{\mathrm{pre}}$, into $V_{\mathrm{post}}$. \sys starts from a healthy modern base $H$, applies a task patch $D$ to obtain task state $C$, and records a restoration patch $G$ that yields restored state $H'$. We require $H'$ to recover the relevant healthy behavior of $H$, not to reproduce it textually. Figure~\ref{fig:overview} summarizes the four stages.

\subsection{Deriving Task Evidence}

\sys derives task evidence from a merged PR in the repository's own history. Its description, implementation patch, and executable checks identify the changed condition, the checks that observe it, and nearby behavior that must remain intact. This stage produces target checks, regression checks, and a source change profile. Target checks expose the condition to be reconstructed; regression checks protect surrounding behavior; the source change profile records the affected components and edit extent.

\rev{For product repositories without prewritten benchmark tasks, \sys derives the statement from a normalized PR title, a title-grounded template, or a concise summary constrained by the PR, commit, and diff. A source-derived target check observes the behavior transferred from the historical change, while a separate protected check covers nearby behavior that must remain stable.}

Candidates are excluded when the changed condition is ambiguous, its checks cannot be executed, or the relevant change cannot be separated from unrelated modifications. The retained evidence provides provenance and the source change profile for later validation.

\subsection{Establishing the Modern Task Base}

For each eligible PR, \sys selects a fixed descendant revision from the same repository lineage as the healthy modern base $H$. It first honors a declared healthy or target revision. If neither is available, it resolves the accessible upstream default branch in a fixed priority order. A fallback to the historical base is recorded separately and is not counted as a modern base. Selection is completed before task construction, and the resolved commit hash is frozen for reproducibility.

The selected revision must support a clean checkout, its runtime dependencies and services, and executable checks. It must also still host the relevant behavior: the mapped target and regression checks must pass on $H$, and there must be a defensible transformation surface in the modern code. Native repository configuration or an external preparation tool may establish this environment; \sys operates after the base is runnable. Appendix~\ref{app:base-selection} gives the full resolution order and acceptance gates.

PR alignment locates the modern realization of the changed behavior, implementation, and checks despite refactoring, file movement, API changes, or test reorganization. A candidate advances only when this correspondence is sufficiently grounded and both target and regression checks pass on $H$.

This stage also instantiates a task specification: the goal presented to the agent, the expected output, the validation oracle, and the permitted edit scope. Our evaluation adapters create an unfixed behavior for Bug Fix, a missing capability for Feature Addition, an observable failure for Test Generation, old API usage for API Migration, or a sandboxed vulnerability for Security Repair. These adapters vary the objective and oracle while sharing one construction and validation core. Additional task families can use the same core whenever their task condition, output contract, and executable oracle can be derived from historical PR evidence on a healthy descendant revision.

\subsection{Constructing the Task State}

\begin{wrapfigure}{r}{.46\linewidth}
\captionsetup{font=footnotesize,skip=4pt,justification=raggedright}
\centering
    \includegraphics[width=\linewidth]{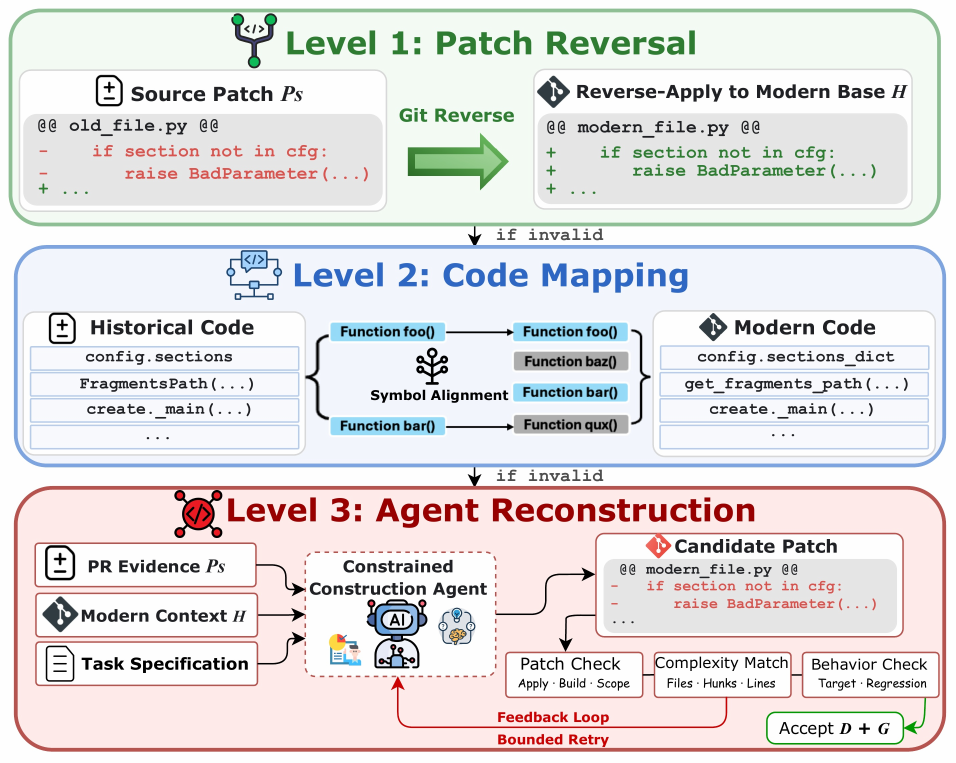}
    \caption{Task state construction in \sys.}
    \label{fig:task-construction}
\end{wrapfigure}

Given the task evidence, healthy modern base, and task specification, \sys constructs $D$ so that $C=\mathrm{apply}(H,D)$ presents the intended maintenance condition while preserving unrelated behavior. Figure~\ref{fig:task-construction} expands the three construction levels. Construction escalates only when an earlier level cannot produce a valid candidate.

\textbf{Level 1: Patch Reversal.} Level 1 reverses the source patch when it remains compatible with the modern context. This route preserves the closest structural link to the change written by the developer.

\textbf{Level 2: Code Mapping.} When direct replay fails but a unique source-block correspondence remains, Level 2 locates the historical post-change block in the modern file. It allows indentation normalization and replaces the block with its reindented pre-change form. The method favors conservative local transformations, reparses the result, and requires the mapped patch to pass the same scope and executable gates. Level 2 makes one structure-guided attempt. Missing, ambiguous, or incompatible correspondence advances the case to Level 3.

\textbf{Level 3: Agent Reconstruction.} When the historical behavior remains present but structural correspondence is insufficient, a construction agent receives PR evidence, relevant modern context, the task specification, and feedback from prior attempts. It proposes scoped candidate patches and filters them through apply, syntax, and scope checks, then ranks survivors by source change profile fidelity and edit complexity before lifecycle validation. A failed candidate returns structured evidence identifying an unmet target condition, broken regression check, restoration failure, or fidelity deviation. The loop allows at most four attempts and terminates at acceptance or budget exhaustion. Cases without a defensible modern behavior host are rejected, and downstream agent outcomes never select or revise tasks. Appendix~\ref{app:l2-l3} gives Level 2 pseudocode and the complete Level 3 protocol.

Patch checks reject candidates that fail to apply, exceed the permitted scope, or break basic syntactic and build integrity. Complexity matching compares their files, hunks, changed lines, symbols, and check surface with the source change profile. For an accepted task patch $D$, \sys records a restoration patch $G$: it is derived as the inverse transformation when possible, otherwise constructed as the corresponding modern repair, and must restore the intended behavior from $C$ to $H'$.

\subsection{Lifecycle and Scope Validation}

\sys validates each candidate over three executable states. On healthy modern base $H$, target and regression checks pass. On task state $C$, target checks expose the specified task condition while regression checks continue to pass. After applying $G$, both check sets pass on restored state $H'$. The target checks therefore pass, fail, and pass across the three states, while regression checks pass throughout. Repeated runs reject unstable cases.

The scope gate permits only artifacts allowed by the task specification and prevents protected checks or metadata from manufacturing validity. The fidelity gate compares the source PR with the modern realization using affected components, edit extent, and check surface. It rejects scope collapse, in which a substantial source change becomes trivial, and scope inflation, in which a narrow change requires broad unrelated edits.

An accepted variant contains $H$, $D$, $G$, the target and regression checks, the task specification, and source PR provenance. Only variants that pass lifecycle, scope, and fidelity validation are frozen for evaluation. The task families therefore demonstrate different agent goals and scoring rules over one reusable construction core.

%% file: body/Experiment.tex
\section{Experiments}\label{sec:eval}

\subsection{Experiment Setup}

\paragraph{Paired task corpus.}
The construction study begins with \matchedAttempts{} \emph{construction-eligible source changes}: cases for which source PR evidence, executable checks, and same-repository base prerequisites are available. The process yields 900 paired sets: 500 Bug Fix sets and 100 each for Feature Addition, Test Generation, Application Programming Interface (API) Migration, and Security Repair. We allocate more Bug Fix sets because repository-level repair has the broadest standardized pool of executable tasks. Every set contains two provenance-linked branches. The Original Branch is the official benchmark case at its historical revision; the \sys Branch reconstructs that case on a healthy modern revision of the same repository. Thus, each modern task has a specific historical counterpart established before agent evaluation. Appendix~\ref{app:candidate-funnel} reports the upstream evidence and readiness funnel.

\paragraph{Tasks and success criteria.}

Each adapter defines a distinct output and verification contract:
\begin{enumerate}
  \item \textbf{Bug Fix.} Repair faulty behavior and pass target and regression checks.
  \item \textbf{Feature Addition.} Implement the requested missing behavior and pass feature and regression checks.
  \item \textbf{Test Generation.} Add a test that fails on the task state, passes after restoration, and respects the test-only scope.
  \item \textbf{API Migration.} Replace obsolete source-API usage with the required target API and pass migration and regression checks.
  \item \textbf{Security Repair.} Remove the vulnerability exposed by a deterministic sandboxed oracle and pass security and regression checks.
\end{enumerate}
A solution is counted as solved only when its family-specific oracle, regression checks, and scope gate all pass.

\paragraph{Construction and evaluation agents.}
Agent Reconstruction uses Claude Code with Opus 4.8 in a bounded loop that generates, executes, and refines candidates. Evaluation uses Codex CLI with GPT-5.5, Claude Code with Sonnet 5, Gemini CLI with Gemini 3.1 Pro, and GitHub Copilot with GPT-5.6 Terra. Each agent receives one clean run on each branch. Within a pair, task intent, model, interface, visible evidence, permissions, context policy, budget, and verifier strength are held constant. Evaluation agents cannot access construction traces, task patches, restoration patches, or hidden target checks. Appendix~\ref{app:agent-protocol} gives the execution limits and complete protocol.

\begingroup
\subsubsection{Ablation Study Setup}
Four controlled studies align with RQ1--RQ4. They measure the contributions of the construction stages, compare semantic assessments across construction levels and reviewer sources, test branch-level outcome consistency under repeated runs, and examine coverage and resource use across construction methods.
\endgroup

\subsection{RQ1: How Reliably Does \sys Reconstruct Tasks from Repository History?}

We collect \matchedAttempts{} construction-eligible changes from 12 public benchmark collections and releases~\cite{jimenez2023swe,zhang2025swe,deng2025swe,li2025feabench,badertdinov2025swe,wang2025swebenchplusplus,islam2023pymigbench,wei2025patcheval,bui2022vul4j,wu2023vjbench}. We define \emph{verified task recovery} as the fraction reaching the finalized corpus after lifecycle, scope, fidelity, and semantic qualification. Appendix~\ref{app:rq1-details} reports exact counts and Wilson 95\% confidence intervals (CIs).

\begin{figure}[htbp]
  \centering
  \includegraphics[width=1.0\linewidth]{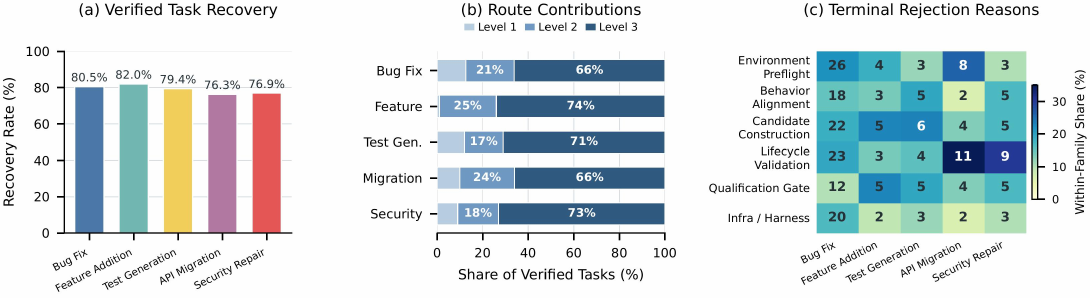}
  \caption{RQ1 recovery, construction routes, and terminal rejection reasons.}
  \label{fig:rq1-results}
\end{figure}

Figure~\ref{fig:rq1-results}(a) shows that \sys reconstructs \fullRecovered{} of \matchedAttempts{} eligible changes, yielding \fullRecoveryRate{} overall recovery. Family estimates range from 76.3\% for API Migration to 82.0\% for Feature Addition, with wider CIs for the four smaller families.

Figure~\ref{fig:rq1-results}(b) assigns each finalized task to its first successful construction level. Level 1 contributes 95 tasks directly; Level 2 contributes 190 after Level 1 fails; Level 3 contributes 615 after both earlier levels are exhausted. The shares are 10.6\%, 21.1\%, and 68.3\%, quantifying the incremental coverage supplied by each level.

Figure~\ref{fig:rq1-results}(c) assigns each of the 230 rejected candidates one terminal reason. Lifecycle validation accounts for 50 cases, environment or preflight failures for 44, and exhausted candidate construction for 42. Attrition is therefore distributed across base readiness, task construction, executable validation, and final qualification.

\Needspace{10\baselineskip}
\begin{wraptable}[6]{r}{.52\linewidth}
\vspace{-2pt}
\captionsetup{font=footnotesize,skip=4pt,justification=raggedright}
\centering
  \small
  \footnotesize\setlength{\tabcolsep}{2pt}
  \begin{tabular}{lrrc}
    \toprule
    \textbf{Method} & \textbf{Verified} & \textbf{Recovery} & \textbf{95\% CI} \\
    \midrule
    Direct reversal & 81/\bugBaselineAttempts{} & \exactRecoveryRate{} & [10.6, 15.9] \\
    SWE-smith PR Mirror & \oneShotRecovered{}/\bugBaselineAttempts{} & \oneShotRecoveryRate{} & [58.4, 66.0] \\
    \sys & 500/\bugBaselineAttempts{} & 80.5\% & [77.2, 83.4] \\
    \bottomrule
  \end{tabular}
  \caption{Matched recovery on \bugBaselineAttempts{} Bug Fix candidates.}
  \label{tab:rq1-baselines}

\end{wraptable}
Table~\ref{tab:rq1-baselines} compares methods on the same 621 Bug Fix candidates under one construction and verification envelope. Direct reversal recovers 81 tasks and SWE-smith PR Mirror 387~\cite{yang2025swesmith}; \sys recovers 500. It therefore adds 113 verified tasks, a 29.2\% relative and 18.2-point absolute gain. Appendix~\ref{app:baseline-protocol} specifies the matched model, evidence, context, attempts, tools, and budgets.

\begingroup
\subsubsection{RQ1 Ablation: Construction Cascade}
On a separate balanced cohort with 100 cases per family, Level 1 alone verifies 42/500 tasks (8.4\%). Adding Level 2 raises recovery to 125/500 (25.0\%), and allowing one Level 3 attempt reaches 330/500 (66.0\%). The full cascade verifies 395/500 (79.0\%): feedback-guided Level 3 attempts recover 65 additional cases after the first attempt. Thus, the adaptive stages account for most verified cases, matching the route distribution in Figure~\ref{fig:rq1-results}(b). A separate 200-case cohort, selected after Levels 1 and 2 fail and Level 3 applicability is confirmed, yields mean one-shot acceptance of 13.0\%, 8.0\%, and 10.0\% for GPT-5.6 Sol, GPT-5.6 Terra, and Gemini 3.7 Flash, respectively. The one-shot results characterize the harder Level 3-eligible cases; the balanced-cohort gains show the contribution of the complete cascade. Appendix~\ref{app:rq1-ablation} visualizes the cascade and all three model repetitions.
\endgroup

\begin{findingbox}{Finding 1} \sys transforms repository history from a passive record into a recoverable source of verified coding tasks through progressively adaptive construction.
\end{findingbox}

\subsection{RQ2: How Faithfully Does \sys Preserve Historical Changes?}

\begin{figure}[htbp]
  \centering

  \includegraphics[width=1.0\linewidth]{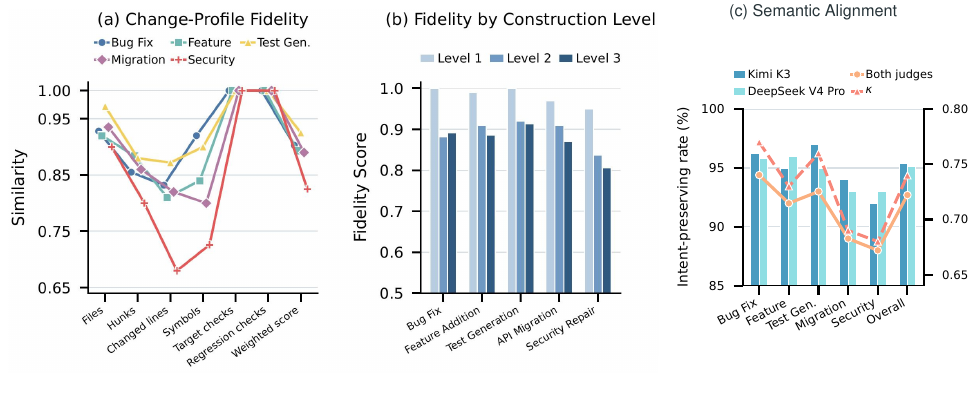}
  \caption{Source-change fidelity and semantic alignment across task families.}
  \label{fig:rq2-results}\label{fig:semantic-audit}
\end{figure}

RQ2 examines how well the 900 finalized tasks retain the complexity, scope, and maintenance meaning of their source changes. We compare each historical source patch with its modern restoration patch along six dimensions: changed files, hunks, lines, symbols, target checks, and regression checks. Dimension similarity is the ratio between smaller and larger counts, with identical empty counts assigned 1. Their weighted combination emphasizes changed lines, followed by hunks, files, the two check surfaces, and symbols.\footnote{Weights allocate 76\% to implementation footprint and 24\% to verification surface. Appendix~\ref{app:rq2-details} reports component values and an equal-weight check.}

\rev{Figure~\ref{fig:rq2-results} (left) reports family means from 0.825 to 0.925,} with a macro average of 0.887 and task-weighted average of 0.894. Test Generation scores 0.925, Bug Fix 0.903, Feature Addition 0.893, API Migration 0.890, and Security Repair 0.825. \rev{The connected profiles show how file, hunk, line, symbol, and check correspondence shape the score.}

\rev{Figure~\ref{fig:rq2-results} (center) reports mean fidelity of 0.992} for Patch Reversal, 0.888 for Code Mapping, and 0.881 for Agent Reconstruction. \rev{Fidelity remains high across all three construction routes, with variation in structural correspondence.}

Kimi K3 and DeepSeek V4 Pro independently review 912 anonymized historical and modern pairs without seeing the construction route, trace, or model. The review records behavioral triggers, expected behavior, affected APIs, source-to-modern mappings, contrastive checks, and restoration. Both judges directly accept 834 pairs. Manual adjudication retains 66 of the remaining 78 and excludes 12, producing 900 tasks. \rev{The right panel of Figure~\ref{fig:semantic-audit} reports each judge's rate, direct joint acceptance, and inter-judge agreement;} 92.7\% of the final corpus is directly accepted by both judges. Appendix~\ref{app:semantic-audit} gives the labels and adjudication protocol.

\begingroup
\subsubsection{RQ2 Ablation: Construction Level and Reviewer Source}
We additionally blind-review 150 pre-qualification construction records, sampling ten records from every family-by-level stratum of the frozen 500-case cohort. The sample contains 150 case-level records and 131 unique cases because 17 cases appear at multiple construction levels. Two company-model judges agree on 130/150 labels (86.7\%, $\kappa=0.744$), while two software engineers agree on 126/150 (84.0\%, $\kappa=0.713$). Both model judges assign a positive intent-preservation label to 139 records, compared with 133 for both engineers. Comparable agreement from software engineers supports the semantic qualification used for the 900 finalized tasks across construction levels. Appendix~\ref{app:rq2-blind} gives the sampling frame and dimension-level results.
\endgroup

The audit also defines the applicability boundary. An adapter needs a historical PR with identifiable intent, a healthy descendant hosting the behavior, an executable oracle, and a bounded edit policy. Outputs may include code, tests, configuration, or other artifacts; grounded evidence and executable validation matter more than a fixed task taxonomy.

\begin{findingbox}{Finding 2} \sys preserves the developer-grounded maintenance contract of historical changes while translating their implementation footprint to evolved code.
\end{findingbox}

\subsection{RQ3: How Consistent Are Agent Outcomes Across the Two Branches?}

The Original Branch is the historical behavioral anchor, and the \sys Branch realizes the same maintenance intent on modern code. Table~\ref{tab:rq3-results} reports branch solve rates, directional differences, complete paired outcomes, raw agreement, and Cohen's \(\kappa\). Agreement is the fraction with the same solved or unsolved outcome; \(\kappa\) adjusts for the branches' marginal solve rates.
\begin{table}[htbp]
  \centering

  \footnotesize
  \renewcommand{\arraystretch}{1.10}
  \setlength{\tabcolsep}{1.5pt}
  \begin{tabularx}{\linewidth}{l l *{8}{>{\centering\arraybackslash}X}}
    \toprule
    \textbf{Task family} & \textbf{Agent / Model} & \textbf{Orig.} & \textbf{C2T} & \textbf{$\Delta$ pp} & \textbf{Both} & \textbf{O only} & \textbf{C only} & \textbf{Agree.} & \textbf{$\kappa$} \\
    \midrule
\rowcolor{black!7}\cellcolor{white} & \cellcolor{white}Codex / GPT-5.5 & 41.0 & 42.2 & \updelta{1.2} & 185 & 20 & 26 & 90.8 & 0.811 \\
\rowcolor{black!10}\cellcolor{white} & \cellcolor{white}Claude / Sonnet 5 & 47.0 & 48.4 & \updelta{1.4} & 210 & 25 & 32 & 88.6 & 0.772 \\
\rowcolor{black!4}\cellcolor{white} & \cellcolor{white}Gemini / Gemini 3.1 Pro & 24.0 & 22.0 & \downdelta{2.0} & 102 & 18 & 8 & 94.8 & 0.853 \\
\rowcolor{black!13}\cellcolor{white}\multirow{-4}{*}{\textbf{Bug Fix}} & \cellcolor{white}Copilot / GPT-5.6 Terra & \textbf{54.0} & \textbf{53.0} & \downdelta{1.0} & 240 & 30 & 25 & 89.0 & 0.779 \\
    \midrule
\rowcolor{black!7}\cellcolor{white} & \cellcolor{white}Codex / GPT-5.5 & 46.0 & 44.0 & \downdelta{2.0} & 39 & 7 & 5 & 88.0 & 0.758 \\
\rowcolor{black!13}\cellcolor{white} & \cellcolor{white}Claude / Sonnet 5 & \textbf{66.0} & \textbf{68.0} & \updelta{2.0} & 58 & 8 & 10 & 82.0 & 0.593 \\
\rowcolor{black!4}\cellcolor{white} & \cellcolor{white}Gemini / Gemini 3.1 Pro & 20.0 & 19.0 & \downdelta{1.0} & 17 & 3 & 2 & 95.0 & 0.841 \\
\rowcolor{black!10}\cellcolor{white}\multirow{-4}{*}{\textbf{Feature Addition}} & \cellcolor{white}Copilot / GPT-5.6 Terra & 50.0 & 54.0 & \updelta{4.0} & 43 & 7 & 11 & 82.0 & 0.640 \\
    \midrule
\rowcolor{black!13}\cellcolor{white} & \cellcolor{white}Codex / GPT-5.5 & \textbf{55.0} & \textbf{57.0} & \updelta{2.0} & 50 & 5 & 7 & 88.0 & 0.757 \\
\rowcolor{black!7}\cellcolor{white} & \cellcolor{white}Claude / Sonnet 5 & 36.0 & 40.0 & \updelta{4.0} & 31 & 5 & 9 & 86.0 & 0.703 \\
\rowcolor{black!4}\cellcolor{white} & \cellcolor{white}Gemini / Gemini 3.1 Pro & 24.0 & 20.0 & \downdelta{4.0} & 19 & 5 & 1 & 94.0 & 0.826 \\
\rowcolor{black!10}\cellcolor{white}\multirow{-4}{*}{\textbf{Test Generation}} & \cellcolor{white}Copilot / GPT-5.6 Terra & 48.0 & 50.0 & \updelta{2.0} & 42 & 6 & 8 & 86.0 & 0.720 \\
    \midrule
\rowcolor{black!13}\cellcolor{white} & \cellcolor{white}Codex / GPT-5.5 & \textbf{60.0} & \textbf{55.0} & \downdelta{5.0} & 52 & 8 & 3 & 89.0 & 0.776 \\
\rowcolor{black!10}\cellcolor{white} & \cellcolor{white}Claude / Sonnet 5 & 45.0 & 48.0 & \updelta{3.0} & 39 & 6 & 9 & 85.0 & 0.699 \\
\rowcolor{black!4}\cellcolor{white} & \cellcolor{white}Gemini / Gemini 3.1 Pro & 12.0 & 11.0 & \downdelta{1.0} & 10 & 2 & 1 & 97.0 & 0.853 \\
\rowcolor{black!7}\cellcolor{white}\multirow{-4}{*}{\textbf{API Migration}} & \cellcolor{white}Copilot / GPT-5.6 Terra & 42.0 & 47.0 & \updelta{5.0} & 36 & 6 & 11 & 83.0 & 0.657 \\
    \midrule
\rowcolor{black!7}\cellcolor{white} & \cellcolor{white}Codex / GPT-5.5 & 35.0 & 30.0 & \downdelta{5.0} & 28 & 7 & 2 & 91.0 & 0.795 \\
\rowcolor{black!13}\cellcolor{white} & \cellcolor{white}Claude / Sonnet 5 & \textbf{55.0} & \textbf{51.0} & \downdelta{4.0} & 46 & 9 & 5 & 86.0 & 0.719 \\
\rowcolor{black!4}\cellcolor{white} & \cellcolor{white}Gemini / Gemini 3.1 Pro & 10.0 & 8.0 & \downdelta{2.0} & 8 & 2 & 0 & 98.0 & 0.878 \\
\rowcolor{black!10}\cellcolor{white}\multirow{-4}{*}{\textbf{Security Repair}} & \cellcolor{white}Copilot / GPT-5.6 Terra & 44.0 & 48.0 & \updelta{4.0} & 37 & 7 & 11 & 82.0 & 0.638 \\
    \bottomrule
  \end{tabularx}
  \caption{Matched outcomes between Original and \sys branches. Orig., C2T, and Agree. are percentages; $\Delta$ is the solve-rate difference in percentage points. Both, O only, and C only count cases solved by both branches, Original only, and \sys only, respectively. Each Bug Fix row has 500 cases and each other row 100. Gray intensity indicates relative solve performance within each task family. Agent labels abbreviate Codex CLI, Claude Code, Gemini CLI, and GitHub Copilot.}
  \label{tab:rq3-results}
\end{table}

Across 3,600 matched agent-task pairs, each branch contains 1,478 solved outcomes, or 41.1\%. Discordance is exactly balanced: 186 outcomes are solved only on the Original Branch and 186 only on the \sys Branch. Aggregate agreement is 89.7\%, Cohen's \(\kappa\) is 0.787, positive agreement is 87.4\%, and solved-set Jaccard overlap is 77.6\%. The paired solve-rate difference is 0.0 points with a 95\% interval of [-1.1, 1.1], and exact McNemar \(p=1.0\). Appendix~\ref{app:rq3-details} defines these statistics.

Agent-level behavior is stable. \rev{Agreement ranges from 86.4\% for GitHub Copilot with GPT-5.6 Terra to 95.3\% for Gemini CLI with Gemini 3.1 Pro}, with aggregate \(\kappa\) from 0.729 to 0.852. Gemini 3.1 Pro combines the highest agreement with lower solve rates. Copilot with GPT-5.6 Terra and Claude Code Sonnet 5 have the highest overall solve rates, while Codex with GPT-5.5 performs best on Test Generation and API Migration.

Directional changes remain small and dispersed. \rev{Claude Code with Sonnet 5 and GitHub Copilot with GPT-5.6 Terra gain 12 and 10 solved tasks} on the \sys Branch, Codex with GPT-5.5 changes by four, and Gemini 3.1 Pro loses 18. No task-family shift exceeds five percentage points, and 11 of 20 comparisons differ by at most two points. \rev{The paired outcomes remain balanced across branches and task families.}

\begingroup
\subsubsection{RQ3 Ablation: Branch and Repetition Sensitivity}
A repeated-run study fixes 200 pair-valid cases, with 40 from each family, and changes only the Original versus \sys branch. Four agent configurations receive three independent runs per branch, producing 4,800 strict-valid evaluations and 2,400 matched outcomes. The two branches solve 975 and 977 outcomes, respectively; 842 pairs are solved on both, 133 only on the Original branch, 135 only on the \sys branch, and 1,290 on neither. Agreement is 88.8\% ($\kappa=0.769$), the paired difference is +0.083 percentage points with a 95\% interval of [-1.254, 1.420], and exact McNemar $p=0.951$. Agent ranking is identical on both branches within every family. The repeated runs reproduce the branch balance and agent ordering seen in the 3,600-pair evaluation, strengthening the conclusion that the modern branch preserves comparative agent outcomes. Appendix~\ref{app:rq3-ablation} reports all family-agent cells and replicate stability.
\endgroup

\begin{findingbox}{Finding 3} \sys-generated tasks retain the comparative evaluation signal of their historical anchors across heterogeneous coding agents.
\end{findingbox}

\subsection{RQ4: How Effectively Does \sys Reuse Executable Repository Environments?}

RQ4 measures the systems benefit of deriving multiple finalized tasks from a healthy modern base. The matched baseline prepares and retains one environment for each of 900 tasks. \rev{\sys prepares 388 healthy modern bases and reuses them across associated variants, including task-specific materialization and verification in the accounting.} Both conditions use the same hardware, retention period, and accounting rules.

\par\addvspace{6pt}
\sbox{\rqfourtable}{%
  \footnotesize\setlength{\tabcolsep}{2pt}%
  \renewcommand{\arraystretch}{1.25}%
  \begin{tabular}{@{}lrrr@{}}
    \toprule
    \textbf{Metric} & \textbf{\shortstack{One Base\\per Task}} & \textbf{\shortstack{Shared Modern\\Bases}} & \textbf{Effect} \\
    \midrule
    Bases prepared & 900 & 388 & 56.9\% fewer \\
    Tasks per base & 1.00 & 2.32 & 2.32$\times$ \\
    Total setup time & 3,240 h & 1,349 h & 58.4\% less \\
    Setup time per task & 3.60 h & 1.50 h & 58.4\% less \\
    Total retained storage & 5,580 GB & 1,607 GB & 71.2\% less \\
    Storage per task & 6.20 GB & 1.79 GB & 71.2\% less \\
    \bottomrule
  \end{tabular}%
}
\noindent
\begin{evidencecolumn}{.59}
  \centering
  \usebox{\rqfourtable}
  \sidecaption{table}{Environment resources with separate and shared modern bases.}
  \label{tab:rq4-amortization}
\end{evidencecolumn}\hfill%
\begin{evidencecolumn}{.38}
  \centering
  \includegraphics[height=\dimexpr\ht\rqfourtable+\dp\rqfourtable\relax]{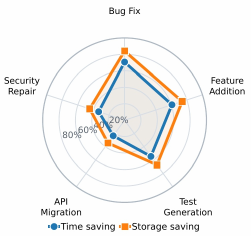}
  \sidecaption{figure}{Family-level time and storage savings from modern-base reuse.}
  \label{fig:rq4-amortization}
\end{evidencecolumn}
\par\addvspace{6pt}

Table~\ref{tab:rq4-amortization} contrasts rebuilding an environment for every task with reusing 388 healthy modern bases. Reuse lowers amortized environment time from \historicalEnvironmentTime{} to \sharedEnvironmentTime{} and storage from \historicalEnvironmentStorage{} to \sharedEnvironmentStorage{} per task. Figure~\ref{fig:rq4-amortization} reports family-level reductions. Bug Fix achieves the greatest reuse and savings; API Migration currently maps each task to a distinct base and benefits less.

Benefits depend on reuse density: families placing several tasks on one maintained base save more than those using distinct bases. Matched hardware, prices, and retention isolate this reuse effect, while RepoLaunch shows how modern reconstruction and historical preparation combine when either route alone is insufficient.

\rev{The matched accounting comparison includes construction automation, compute, storage, and retention.} Reuse lowers expenditure from \$1,917 to \$1,710. The \$207 saving equals \$0.23 per task and a 10.8\% net reduction, including additional automation and Agent Reconstruction expense. Appendix~\ref{app:rq4-details} gives the accounting breakdown.

RepoLaunch supplies a complementary path when modern reconstruction is unavailable. It makes another 195 historical revisions executable, raising coverage from 900 to 1,095 of 1,130 cases, or 96.9\%, and leaving 35 unresolved. \rev{This route complements modern task construction with historical environment preparation.}

\begingroup
\subsubsection{RQ4 Ablation: Method and Resource Trade-off}
On the same 100 Bug Fix cases, Direct Reversal qualifies 13, PR Mirror qualifies 62, and the full \sys cascade qualifies 81. The cascade gains 19 cases over PR Mirror with 215 model calls, compared with 100 for PR Mirror and none for Direct Reversal. This construction cost is offset at the corpus level: sharing 388 modern bases reduces setup time by 58.4\% and retained storage by 71.2\%, while complete expenditure falls from \$1,917 to \$1,710. The matched comparisons show how broader recovery and base reuse jointly produce the net saving in RQ4. Appendix~\ref{app:rq4-ablation} reports the matched outcomes and primary resource measures.
\endgroup

\begin{findingbox}{Finding 4} \sys turns executable repositories into reusable task infrastructure and complements historical environment preparation when modern reconstruction is unavailable.
\end{findingbox}

\begingroup
\subsection{RQ5: How Does \sys Perform in Industrial Software Development?}\label{sec:product}

\paragraph{Industrial deployment.}
We construct 1,000 tasks from 11 private repositories, each with a distinct historical PR, source patch, and task patch. The corpus covers 355 Bug Fix, 480 Feature Addition, 43 Test Generation, 46 API Migration, and 76 Security Repair cases. Private PR histories require us to recover task intent from titles, commits, and diffs and derive executable target and regression checks from source evidence. We then apply the same modern-base construction and healthy--challenge--restored validation used in the public study. Task statements comprise 727 normalized PR titles, 39 title-grounded templates, and 234 evidence-grounded summaries. This provenance and the five task contracts give the product corpus breadth beyond a single repository or repair type. Appendix~\ref{app:product} reports repository coverage and the qualification protocol.

\paragraph{Canonical task evaluation.}
Every task has executable target and regression checks; the evaluation uses fresh workspaces and hidden checks. Across three independent runs on all 1,000 cases, GPT-5.6 Sol solves 146 tasks at least once, yielding mean Pass@1 of 11.03\% and Pass@3 of 14.6\%. \rev{Pass@1 is the mean solve rate across runs; Pass@3 counts tasks solved in at least one run.} This full-corpus run establishes that the constructed cases can be executed and scored at product scale. Appendix~\ref{app:product} reports wave-level outcomes and selection details.

\subsubsection{Ablation: Cross-Model Comparison}
\Needspace{15\baselineskip}
\begin{wrapfigure}[13]{r}{.46\linewidth}
\vspace{-\intextsep}
\captionsetup{font=footnotesize,skip=4pt,justification=raggedright}
\centering
  \includegraphics[width=\linewidth]{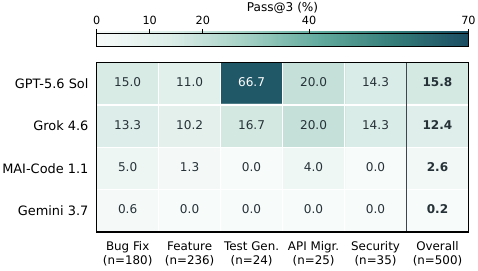}
  \caption{RQ5 Pass@3 by model and task family on the fixed product comparison set. Cell color and labels report the percentage of tasks solved at least once across three runs.}
  \label{fig:product-model}

\end{wrapfigure}
The fixed comparison cohort samples 250 cases from each wave and retains all five families: 180 Bug Fix, 236 Feature Addition, 24 Test Generation, 25 API Migration, and 35 Security Repair. Four models share the agent scaffold, prompt, tools, hidden checks, 12-turn limit, and 1,800-second timeout. Pass@3 is 15.8\% for GPT-5.6 Sol, 12.4\% for Grok 4.6, 2.6\% for MAI-Code 1.1 Flash, and 0.2\% for Gemini 3.7 Flash.

The fixed evaluator separates model outcomes within task families; for Test Generation, GPT-5.6 Sol solves 16/24 cases and Grok 4.6 solves 4/24. Together with the full-corpus run, this comparison shows that source-grounded product tasks support repeatable construction, executable scoring, and downstream model assessment. The deployment exercises the adaptive construction, source-linked qualification, fixed-condition agent evaluation, and modern-base lifecycle examined in RQ1--RQ4. Appendix~\ref{app:product} gives family-level outcomes and verification details.

\begin{findingbox}{Finding 5} \sys turns private PR histories into 1,000 executable tasks across five families and supports discriminating agent evaluation on a fixed product cohort.
\end{findingbox}
\endgroup

%% file: body/Conclusion.tex
\section{Conclusion and Outlook}\label{sec:conclusion}

We introduced \sys, a history-grounded framework that reconstructs merged PRs as executable coding-agent tasks on healthy descendant revisions. Its three adaptive construction routes and lifecycle, scope, and fidelity checks transfer developer-grounded tasks onto maintained code. Across five task families, \sys finalizes 900 of 1,130 eligible changes, improves matched Bug Fix recovery by \recoveryGain{}, and produces tasks with \rev{0.894 task-weighted source change profile fidelity and 89.7\% aggregate matched agent-outcome agreement.} Modern-base reuse also reduces environment time by \environmentTimeSaving{} and storage by \environmentStorageSaving{}. \rev{These savings reduce the resources needed to produce executable agent tasks.}

\rev{The RQ-specific analyses show that adaptive construction, independent semantic review, branch matching, and base reuse each contribute distinct evidence. The industrial study extends the same lifecycle to 1,000 private-history tasks and a fixed cross-model evaluation.}

Future work will extend \sys to more task families, languages, and build systems. We will improve oracle generation as tests and APIs evolve, and support task transfer across forks, successor projects, and related dependency ecosystems through traceable interface mappings. We will also repeat paired evaluations across model releases. Together, these extensions will expand the historical changes that can become current executable data for agent training and continuous evaluation.

%% file: body/Appendix.tex
\section{Experimental Materials and Protocols}\label{app:experimental-protocols}

\subsection{Corpus and Task Provenance}\label{app:eval-details}

\paragraph{Eligibility Funnel.}\label{app:candidate-funnel}

The 1,130 construction-eligible changes follow a fixed readiness funnel:
1,783 raw source records, 1,521 with sufficient evidence, 1,233 with a runnable
modern base, and 1,197 with aligned behavior and checks. Construction verifies
912 candidates; semantic qualification finalizes 900. Eligibility is fixed
before agent evaluation.

\paragraph{Task Provenance.}

Table~\ref{tab:app-sources} reports the provenance of the 900 paired evaluation sets. Each Original Branch is the official case from the listed source, and its \sys Branch is reconstructed from that same case.
The task families include Application Programming Interface (API) Migration.

\begin{table}[H]
  \centering
  \small
  \setlength{\tabcolsep}{1.0mm}
  \footnotesize\begin{evidencecolumn}{.47}\centering
\begin{tabular}{llr}
    \toprule
    \textbf{Task family} & \textbf{Source corpus} & \textbf{Cases} \\
    \midrule
    Bug Fix & SWE-bench Live & 129 \\
    Bug Fix & SWE-bench Lite & 30 \\
    Bug Fix & SWE-Bench Pro & 101 \\
    Bug Fix & SWE-bench Original & 151 \\
    Bug Fix & SWE-bench Verified & 89 \\
    \midrule
    Feature Addition & FEA-Bench & 30 \\
    Feature Addition & SWE-rebench V2 PRs & 68 \\
    Feature Addition & SWE-Bench++ & 2 \\
    \bottomrule
  \end{tabular}\end{evidencecolumn}\hfill%
\begin{evidencecolumn}{.51}\centering
\begin{tabular}{llr}
    \toprule
    \textbf{Task family} & \textbf{Source corpus} & \textbf{Cases} \\
    \midrule
    Test Generation & SWE-bench Live & 20 \\
    Test Generation & SWE-bench Lite & 2 \\
    Test Generation & SWE-bench Verified & 39 \\
    Test Generation & SWE-Bench Pro & 5 \\
    Test Generation & SWE-bench Original & 34 \\
    \midrule
    API Migration & PyMigBench & 100 \\
    \midrule
    Security Repair & PatchEval executable subset & 84 \\
    Security Repair & Vul4J & 15 \\
    Security Repair & VJBench & 1 \\
    \bottomrule
  \end{tabular}\end{evidencecolumn}\par
  \caption{Provenance of the paired evaluation sets.}
  \label{tab:app-sources}
\end{table}

\subsection{Construction and Validation Protocol}\label{app:construction-protocol}

\paragraph{Modern Base Selection.}\label{app:base-selection}

\sys selects a base within the upstream lineage of the source PR. It resolves
a declared healthy revision first, followed by a declared
target revision and the repository's remote default revision. If the default
cannot be resolved, it tries the \emph{main}, \emph{master}, \emph{devel},
\emph{develop}, and \emph{trunk} branches in that order. The historical base
revision is the final fallback. We tag these fallbacks separately and exclude
them from the descendant modern bases used in the environment reuse analysis.

Before freezing the resolved commit hash, \sys requires a clean checkout,
available runtime dependencies and services, collectable mapped checks,
passing target and regression checks on the healthy state, a surviving modern
host for the PR behavior, and a permitted transformation surface. The first
eligible resolution is frozen before construction, preventing post-hoc base
selection.

\paragraph{Code Mapping and Agent Reconstruction.}\label{app:l2-l3}

Level 2: Code Mapping uses a unique source-block correspondence. It extracts
consecutive edit groups from non-test source hunks. Removed lines form the
historical pre-change block, and added lines form the post-change block. Pure
additions, pure deletions, import-only changes, and missing modern files are
outside this route. Algorithm~\ref{alg:level2} summarizes the conservative
configuration used in our study.

\begin{algorithm}[htbp]
\hrule
\caption{Level 2: Code Mapping}
\label{alg:level2}
\hrule
\smallskip
\textbf{Input}: source patch \(P_s\), healthy modern base \(H\)\\
\textbf{Output}: task patch \(D\), or \textsc{Fail}
\begin{enumerate}
  \setlength{\itemsep}{0pt}
  \setlength{\parsep}{0pt}
  \item \(E \leftarrow\) paired pre-change and post-change blocks from
        eligible source hunks.
  \item \textbf{for each} \((b^{-}, b^{+}, p) \in E\) \textbf{do}
  \item \hspace{1em}\(f \leftarrow\) resolve \(p\), then its
        source-directory variant, in \(H\).
  \item \hspace{1em}\textbf{if} \(f\) is unresolved \textbf{then return}
        \textsc{Fail}.
  \item \hspace{1em}\(L \leftarrow\) exact occurrences of \(b^{+}\) in \(f\).
  \item \hspace{1em}\textbf{if} \(|L| \neq 1\) \textbf{then}
        \(L \leftarrow\) indentation-normalized occurrences of \(b^{+}\).
  \item \hspace{1em}\textbf{if} \(|L| \neq 1\) \textbf{then return}
        \textsc{Fail}.
  \item \hspace{1em}Reindent \(b^{-}\) to the unique location in \(L\).
  \item \hspace{1em}Replace \(b^{+}\) with the reindented \(b^{-}\).
  \item \textbf{end for}
  \item Reparse modified files and compute \(D=\operatorname{diff}(H,H')\).
  \item \textbf{if} \(D\) is empty, invalid, or touches a protected artifact
        \textbf{then return} \textsc{Fail}.
  \item \textbf{return} \(D\).
\end{enumerate}
\hrule
\end{algorithm}

The optional abstract syntax tree fallback matches a whole function by an
identical qualified name and reparses the resulting file. The reported
conservative configuration disables this fallback to preserve modern logic
outside the uniquely matched block. Code Mapping handles indentation and local
context drift. Renamed or relocated symbols, split or merged functions,
repeated target blocks, and behavior moved across abstraction layers require
Level 3: Agent Reconstruction.

Agent Reconstruction receives the historical PR evidence, the modern behavior
host and call context, task checks, protected regression checks, the source
change profile, and prior failure evidence. Scoped diff generation handles
bounded code contexts, while full-agent worktree editing supports cases that
require call-chain inspection. Larger changes can enable adaptive
multi-candidate generation within the same attempt budget. Each candidate
first passes non-empty-diff, apply, syntax, permitted-artifact, and basic
fidelity checks. Surviving candidates are ranked by source change profile
fidelity and edit complexity, and the highest-ranked candidate enters full
lifecycle validation. The construction agent receives structured feedback for
application, target-condition, regression, restoration, scope, or fidelity
failures. The loop terminates after acceptance or at most four attempts. For
an accepted task patch \(D\), the restoration patch \(G\) comes from a clean
inverse modern transformation or a scoped modern repair that passes the same
lifecycle and fidelity checks. Cases without a modern behavior host are
rejected.

\paragraph{Matched PR Mirror Protocol.}\label{app:baseline-protocol}

The PR Mirror baseline and \sys operate on the same 621 Bug Fix candidates,
healthy modern bases, source PR evidence, mapped checks, and relevant modern
context. Both use Claude Code with Opus 4.8 under the same total attempt
limit, token and time budgets, tool permissions, and final verifier. Their
prompts share the same evidence scaffold and output constraints. PR Mirror
generates independent one-shot reversal candidates. Each method can generate
at most four candidates. Both receive the same aggregate token and wall-clock
budget across those candidates. \sys can continue with Code Mapping and
structured verifier feedback when direct reversal fails. This matched
protocol holds the candidate pool and execution envelope fixed while
measuring the construction procedures. Bug Fix provides the matched comparison
because its failure-inducing reversal, repair output, and test contract align
with the original PR Mirror task definition.

\paragraph{Fidelity Gate.}

The fidelity gate compares the historical source patch \(P_s\) and modern
restoration patch \(G\) across files, hunks, changed lines, symbols, target
checks, and regression checks. Both patches encode the forward maintenance
change, while task patch \(D\) constructs the task state in the inverse
direction. The six fidelity weights are 0.18, 0.20, 0.28, 0.10, 0.12, and
0.12. The gate requires an aggregate score of at least 0.65, line, hunk, and
file ratios of at least 0.50, a line ratio no greater than 2.50, and a
regression-check ratio of at least 0.25. These thresholds reject localized
simplifications, over-expanded edits, and insufficient regression surfaces.
The equal-weight sensitivity analysis is reported in
Section~\ref{app:rq2-details}.

\subsection{Agent Execution Protocol}\label{app:agent-protocol}

Agent Reconstruction uses Claude Code with Opus 4.8 and bounded execution
feedback. Evaluation uses the Codex command line interface (CLI) with GPT-5.5,
Claude Code with Sonnet 5, Gemini CLI with Gemini 3.1 Pro, and GitHub Copilot
with GPT-5.6 Terra. Each agent, set, and branch receives one authoritative
clean run with a limit of 1,800 seconds. Final verification has a 300-second
limit. Infrastructure failures are rerun separately and do not replace
completed behavioral outcomes.

Within each pair, the two branches use the same model version, interface, task intent, visible evidence, tool permissions, context policy, and budgets. A solution is accepted only when its target and regression checks pass and its changes satisfy the task-specific scope. Evaluation agents cannot access construction prompts or traces, the task patch, the restoration patch, or hidden target checks.

\begingroup
\Needspace{13\baselineskip}
\section{Task Recovery and Construction}\label{app:rq1-details}\label{app:supplementary-results}
\subsection{Supplementary Results}

\par\addvspace{6pt}\noindent
\begin{evidencecolumn}{.47}
\textbf{Recovery and routes.} Table~\ref{tab:app-rq1-recovery} reports eligible and finalized cases, recovery rates, and the first successful construction level. Each accepted task contributes to exactly one route. \textbf{Overall recovery is 900/1,130 (79.6\%); Level 3 contributes 615 of the 900 finalized tasks.}

Family-level Wilson 95\% CIs are [77.2, 83.4] for Bug Fix, [74.2, 87.8] for Feature Addition, [71.5, 85.5] for Test Generation, [68.4, 82.8] for API Migration, and [69.0, 83.3] for Security Repair.\par
\end{evidencecolumn}\hfill%
\begin{evidencecolumn}{.50}
\centering\small\color{black}
  \footnotesize\setlength{\tabcolsep}{2.5pt}
  \begin{tabular}{@{}lrrcc@{}}
    \toprule
    \textbf{Family} & \textbf{Eligible} & \textbf{Final} & \textbf{Recovery} & \textbf{L1/L2/L3} \\
    \midrule
    Bug Fix & 621 & 500 & 80.5\% & 63/106/331 \\
    Feature & 122 & 100 & 82.0\% & 1/25/74 \\
    Test Gen. & 126 & 100 & 79.4\% & 12/17/71 \\
    API Migr. & 131 & 100 & 76.3\% & 10/24/66 \\
    Security & 130 & 100 & 76.9\% & 9/18/73 \\
    \midrule
    \textbf{Overall} & \textbf{1,130} & \textbf{900} & \textbf{79.6\%} & \textbf{95/190/615} \\
    \bottomrule
  \end{tabular}
  \sidecaption{table}{Verified recovery and disjoint construction routes by family. The overall recovery 95\% CI is [77.2, 81.9].}
  \label{tab:app-rq1-recovery}\label{tab:app-rq1-levels}
\end{evidencecolumn}
\par\addvspace{6pt}

\textbf{Terminal rejection.} The 230 rejected candidates comprise 50 lifecycle failures, 44 environment or preflight failures, 42 exhausted constructions, 33 behavior-alignment failures, 31 final qualification failures, and 30 persistent harness failures. These categories are mutually exclusive.

\subsection{Ablation: Construction Cascade and Model Sensitivity}\label{app:rq1-ablation}

\par\addvspace{6pt}\noindent
\begin{evidencecolumn}{.54}
The balanced construction cohort contains 100 cases per family. Cumulative acceptance progresses from 42/500 at Level 1 to 125/500 after Level 2, 330/500 after one Level 3 attempt, and 395/500 with feedback. The added counts are 83, 205, and 65, respectively. The final family counts are 81, 82, 79, 76, and 77 in the order used in Table~\ref{tab:app-rq1-recovery}; feedback attempts two through four contribute 32, 20, and 13 cases.

The separate sensitivity cohort contains 200 Level 3 eligible cases after Levels 1 and 2 fail: 41 Bug Fix, 37 Feature Addition, 41 Test Generation, 41 API Migration, and 40 Security Repair. Three independent one-shot runs accept 24, 26, and 28 cases for GPT-5.6 Sol; 14, 16, and 18 for GPT-5.6 Terra; and 19, 21, and 20 for Gemini 3.7 Flash. \textbf{These are conditional one-shot Level 3 rates on the hard cohort.}
\end{evidencecolumn}\hfill%
\begin{evidencecolumn}{.43}
\centering

  \includegraphics[width=\linewidth]{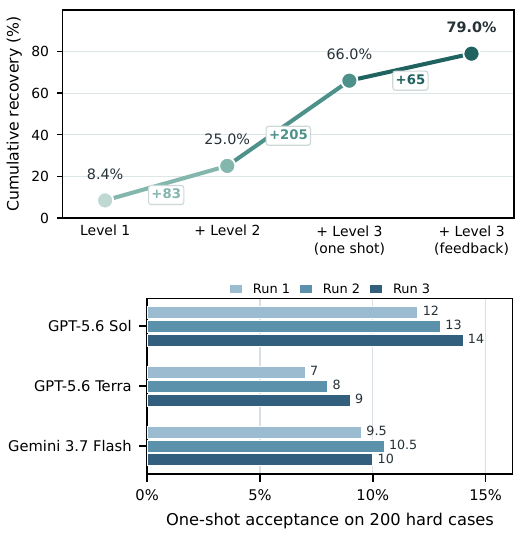}
  \sidecaption{figure}{Construction-stage recovery on the balanced 500-case cohort (top) and one-shot acceptance across three runs on a separate 200-case Level 3-eligible cohort (bottom).}
  \label{fig:app-rq1-cascade-sensitivity}
\end{evidencecolumn}
\par\addvspace{6pt}

\endgroup

\begingroup
\Needspace{12\baselineskip}
\section{Fidelity and Semantic Review}\label{app:rq2-details}
\subsection{Supplementary Results}

\Needspace{13\baselineskip}
\begin{wraptable}[10]{r}{.50\linewidth}
\vspace{-\intextsep}
\captionsetup{font=footnotesize,skip=4pt,justification=raggedright}
\centering\small\color{black}
  \footnotesize\setlength{\tabcolsep}{2.7pt}
  \begin{tabular}{@{}lrrrrr@{}}
    \toprule
    \textbf{Family} & \textbf{Files} & \textbf{Hunks} & \textbf{Lines} & \textbf{Symbols} & \textbf{Score} \\
    \midrule
    Bug Fix & .928 & .855 & .832 & .920 & .903 \\
    Feature & .920 & .884 & .810 & .840 & .893 \\
    Test Gen. & .972 & .880 & .872 & .900 & .925 \\
    API Migr. & .935 & .860 & .820 & .800 & .890 \\
    Security & .900 & .800 & .680 & .726 & .825 \\
    \midrule
    Macro mean & .931 & .856 & .803 & .837 & .887 \\
    \bottomrule
  \end{tabular}
  \caption{Mean source-change profile fidelity across 900 finalized tasks. Target and regression check components are 1.000 in every family.}
  \label{tab:app-rq2-components}\label{tab:app-rq2-distributions}

\end{wraptable}
\textbf{Profile fidelity.} Table~\ref{tab:app-rq2-components} gives the implementation component means for the finalized 900-task corpus. The implementation footprint receives 76\% of the weight and the check surface 24\%. The macro mean is .887 and the task-weighted mean is .894. Family medians (interquartile ranges) are .907 [.837, .977], .890 [.816, .958], .928 [.860, .982], .884 [.776, .949], and .812 [.732, .902] in table order. Equal component weights produce family scores of .923, .909, .937, .903, and .851 for Bug Fix, Feature Addition, Test Generation, API Migration, and Security Repair; the family ordering is unchanged.

\textbf{Construction levels.} Mean fidelity is .992 across 95 Level 1 tasks, .888 across 190 Level 2 tasks, and .881 across 615 Level 3 tasks. The route scores trace the change in structural correspondence.

\textbf{Semantic audit.}\label{app:semantic-audit} Kimi K3 and DeepSeek V4 Pro independently reviewed 912 anonymized candidates without the construction route or outcome. Both directly accepted 834; manual adjudication accepted 66 of the remaining 78 and excluded 12, leaving 900 finalized tasks. \textbf{Both judges directly accepted 92.7\% of the final corpus.} Review covered triggers, expected behavior, invariants, mappings, contrastive checks, and restoration.

\subsection{Ablation: Reviewer-Source Comparison}\label{app:rq2-blind}

The blind frame samples ten distinct case-level records in each family-by-level stratum of the frozen 500-case construction attempts before final qualification. Its 150 records represent 131 cases; 15 cases appear at two levels and two at three. Company-model judges agree on 130/150 labels (86.7\%, $\kappa=.744$); software engineers agree on 126/150 (84.0\%, $\kappa=.713$). Both model judges give a positive intent-preservation label to 139 records, compared with 133 for both engineers. Human agreement is 90.0\% for triggers, 88.0\% for expected behavior, 86.7\% for invariants, 89.3\% for source-to-modern mapping, and 74.7\% for contrastive evidence. \textbf{The two reviewer sources produce comparable intent-label agreement.}
\endgroup

\begingroup
\section{Paired Outcome Consistency}\label{app:rq3-details}
\subsection{Supplementary Results}

Across 3,600 original-versus-modern pairs, the contingency counts are 1,292 both solved, 186 Original only, 186 \sys only, and 1,936 neither. Raw agreement is 89.7\%, Cohen's $\kappa=.787$, positive agreement is 87.4\%, negative agreement is 91.2\%, and solved-set Jaccard overlap is 77.6\%. The paired solve-rate difference is 0.0 percentage points (95\% CI $[-1.1,1.1]$); exact McNemar $p=1.0$. \textbf{Discordant outcomes are balanced.}

\subsection{Ablation: Repeated-Run Branch Consistency}\label{app:rq3-ablation}

The repeated-run study uses 200 pair-valid cases (40 per family), four agents, and three independent runs per branch: 4,800 strict-valid evaluations and 2,400 matched outcomes. Both-solved, Original-only, \sys-only, and neither counts are 842, 133, 135, and 1,290. Original and \sys solve 975 and 977 outcomes; agreement is 88.8\% ($\kappa=.769$), the paired difference is +0.083 points (95\% CI $[-1.254,1.420]$), and exact McNemar $p=.951$. \textbf{Agent ranking is identical across branches within every family.}

\begin{table}[htbp]

  \centering\footnotesize\color{black}
  \renewcommand{\arraystretch}{0.99}
  \setlength{\tabcolsep}{3pt}
  \begin{tabularx}{\linewidth}{l l l >{\hsize=.9\hsize\centering\arraybackslash}X >{\hsize=1.2\hsize\centering\arraybackslash}X >{\hsize=1.2\hsize\centering\arraybackslash}X >{\hsize=.7\hsize\centering\arraybackslash}X}
    \toprule
    \textbf{Task family} & \textbf{Agent} & \textbf{Model} & \textbf{Original} & \textbf{Change2Task} & \textbf{Agreement} & \textbf{$\kappa$} \\
    \midrule
    \rowcolor{black!8} & Codex CLI & GPT-5.6 Terra & 49/120 & 50/120 & 90.8\% & .811 \\
    \rowcolor{black!8} & Claude Code & GPT-5.6 Sol & 56/120 & 58/120 & 88.3\% & .766 \\
    \rowcolor{black!8} & Gemini CLI & Gemini 3.7 Flash & 29/120 & 27/120 & \textbf{95.0\%} & .860 \\
    \rowcolor{black!8}\multirow{-4}{*}{\textbf{Bug Fix}} & GitHub Copilot & GPT-5.6 Terra & 65/120 & 64/120 & 89.2\% & .782 \\
    \midrule
     & Codex CLI & GPT-5.6 Terra & 55/120 & 53/120 & 88.3\% & .764 \\
     & Claude Code & GPT-5.6 Sol & 79/120 & 81/120 & 81.7\% & .588 \\
     & Gemini CLI & Gemini 3.7 Flash & 24/120 & 22/120 & \textbf{95.0\%} & .839 \\
    \multirow{-4}{*}{\textbf{Feature Addition}} & GitHub Copilot & GPT-5.6 Terra & 60/120 & 65/120 & 82.5\% & .650 \\
    \midrule
    \rowcolor{black!8} & Codex CLI & GPT-5.6 Terra & 66/120 & 68/120 & 88.3\% & .764 \\
    \rowcolor{black!8} & Claude Code & GPT-5.6 Sol & 43/120 & 48/120 & 85.8\% & .700 \\
    \rowcolor{black!8} & Gemini CLI & Gemini 3.7 Flash & 29/120 & 24/120 & \textbf{94.2\%} & .831 \\
    \rowcolor{black!8}\multirow{-4}{*}{\textbf{Test Generation}} & GitHub Copilot & GPT-5.6 Terra & 57/120 & 60/120 & 85.8\% & .717 \\
    \midrule
     & Codex CLI & GPT-5.6 Terra & 72/120 & 66/120 & 88.3\% & .762 \\
     & Claude Code & GPT-5.6 Sol & 54/120 & 58/120 & 85.0\% & .699 \\
     & Gemini CLI & Gemini 3.7 Flash & 14/120 & 13/120 & \textbf{97.5\%} & .875 \\
    \multirow{-4}{*}{\textbf{API Migration}} & GitHub Copilot & GPT-5.6 Terra & 50/120 & 56/120 & 83.3\% & .663 \\
    \midrule
    \rowcolor{black!8} & Codex CLI & GPT-5.6 Terra & 42/120 & 36/120 & 91.7\% & .811 \\
    \rowcolor{black!8} & Claude Code & GPT-5.6 Sol & 66/120 & 61/120 & 85.8\% & .716 \\
    \rowcolor{black!8} & Gemini CLI & Gemini 3.7 Flash & 12/120 & 10/120 & \textbf{98.3\%} & .900 \\
    \rowcolor{black!8}\multirow{-4}{*}{\textbf{Security Repair}} & GitHub Copilot & GPT-5.6 Terra & 53/120 & 57/120 & 81.7\% & .631 \\
    \bottomrule
  \end{tabularx}
  \caption{Repeated-run paired outcomes by task family, agent, and model. Each row contains 120 matched branch outcomes; bold marks the highest agreement in each family.}
  \label{tab:app-rq3-ablation-summary}\label{tab:app-rq3-ablation-cells}
\end{table}
\endgroup

\begingroup
\section{Environment Reuse and Cost}\label{app:rq4-details}
\subsection{Supplementary Results}

\textbf{Resource accounting.} Table~\ref{tab:app-rq4-family} reports the per-family resource use and the matched 900-task cost comparison. A base includes checkout, dependencies, services, checks, and healthy-state preflight; retained storage counts shared layers once. Compute costs $\$0.3333$ per hour and storage plus registry retention $\$0.0484$ per GB at the study boundary. Costs use unrounded measurements and include construction attempts. \textbf{Shared bases save $\$207$ across 900 finalized tasks.}

\begin{table}[H]

  \centering\small\color{black}
  \setlength{\tabcolsep}{2.4pt}
  \begin{evidencecolumn}{.52}
\centering\textbf{Panel A: Shared-base resource use}\par\smallskip
  \begin{tabular}{@{}lrrrr@{}}
    \toprule
    \textbf{Family} & \textbf{Tasks} & \textbf{Bases} & \textbf{h/task} & \textbf{GB/task} \\
    \midrule
    Bug Fix & 500 & 116 & 1.06 & 0.99 \\
    Feature & 100 & 39 & 1.41 & 1.62 \\
    Test Gen. & 100 & 49 & 1.63 & 2.02 \\
    API Migr. & 100 & 100 & 2.75 & 4.06 \\
    Security & 100 & 84 & 2.40 & 3.42 \\
    \midrule
    Overall & 900 & 388 & 1.50 & 1.79 \\
    \bottomrule
  \end{tabular}\end{evidencecolumn}\hfill%
\begin{evidencecolumn}{.45}\centering\textbf{Panel B: Matched cost for 900 tasks (USD)}\par\smallskip
  \begin{tabular}{@{}lrr@{}}
    \toprule
    \textbf{Component} & \textbf{Separate} & \textbf{Shared} \\
    \midrule
    Compute & 1,080.00 & 449.67 \\
    Storage & 270.00 & 77.76 \\
    Automation/model & 567.00 & 1,182.57 \\
    \midrule
    \textbf{Total} & \textbf{1,917.00} & \textbf{1,710.00} \\
    \bottomrule
  \end{tabular}
  \end{evidencecolumn}\par\caption{Environment resources by family and matched cost. Shared bases save $\$207$ overall, or $\$0.23$ per finalized task.}
  \label{tab:app-rq4-family}\label{tab:app-rq4-cost}
\end{table}

\textbf{Complementarity.} Among 1,130 eligible changes, 680 are executable through both \sys and RepoLaunch, 220 through \sys alone, 195 through RepoLaunch alone, and 35 through neither. Combined coverage reaches 1,095/1,130 (96.9\%).

\subsection{Ablation: Matched Construction Methods}\label{app:rq4-ablation}

On the same 100 Bug Fix cases, Direct Reversal qualifies 13, PR Mirror 62, and the full \sys cascade 81. Model calls are 0, 100, and 215; total input tokens are 0, 1.62 M, and 4.95 M; cached tokens are 0, 45 K, and 59.8 M. Recorded wall time is 285.4, 2,280, and 49,200 seconds, respectively. \textbf{The full cascade adds 19 qualified tasks over PR Mirror on this fixed cohort.} The separate resource comparison above accounts for the shared-base benefit.
\endgroup

\begingroup
\section{Industrial Software Evaluation}\label{app:product}

The deployment applies the same PR-evidence, modern-base, construction, and healthy--challenge--restored validation lifecycle to 11 private repositories. The final 1,000 cases contain 1,000 distinct PRs, source patches, and task patches. Task statements comprise 727 normalized PR titles, 39 title-grounded templates, and 234 summaries constrained by PR, commit, and diff evidence. \textbf{Every case has a source-linked task patch and executable checks.}

Table~\ref{tab:app-product-repos} summarizes repository coverage, task-family mix, and model outcomes.

\begin{table}[htbp]
  \centering

  \footnotesize\color{black}
  \renewcommand{\arraystretch}{1.03}
  \setlength{\tabcolsep}{2pt}
  \textbf{Panel A: Repository coverage}\par\vspace{3pt}
  \begin{tabularx}{\linewidth}{l l *{6}{>{\centering\arraybackslash}X}}
    \toprule
    \textbf{Repository group} & \textbf{ID} & \multicolumn{2}{c}{\textbf{Wave 1}} & \multicolumn{2}{c}{\textbf{Wave 2}} & \multicolumn{2}{c}{\textbf{Canonical cohort}} \\
    \cmidrule(lr){3-4}\cmidrule(lr){5-6}\cmidrule(lr){7-8}
    & & \textbf{$n$} & \textbf{Share (\%)} & \textbf{$n$} & \textbf{Share (\%)} & \textbf{$n$} & \textbf{Share (\%)} \\
    \midrule
    Early & R1 & 15 & 3.0 & 0 & 0.0 & 15 & 1.5 \\
    \rowcolor{black!6}Early & R2 & 31 & 6.2 & 0 & 0.0 & 31 & 3.1 \\
    \rowcolor{black!12}\textbf{Early subtotal} &  & \textbf{46} & \textbf{9.2} & \textbf{0} & \textbf{0.0} & \textbf{46} & \textbf{4.6} \\
    \midrule
    Active & R3 & 36 & 7.2 & 13 & 2.6 & 49 & 4.9 \\
    \rowcolor{black!6}Active & R4 & 20 & 4.0 & 0 & 0.0 & 20 & 2.0 \\
    Active & R5 & 78 & 15.6 & 23 & 4.6 & 101 & 10.1 \\
    \rowcolor{black!6}Active & R6 & 49 & 9.8 & 6 & 1.2 & 55 & 5.5 \\
    Active & R7 & 51 & 10.2 & 26 & 5.2 & 77 & 7.7 \\
    \rowcolor{black!6}Active & R8 & 99 & 19.8 & 90 & 18.0 & 189 & 18.9 \\
    Active & R9 & 100 & 20.0 & 336 & 67.2 & 436 & 43.6 \\
    \rowcolor{black!6}Active & R10 & 4 & 0.8 & 1 & 0.2 & 5 & 0.5 \\
    Active & R11 & 17 & 3.4 & 5 & 1.0 & 22 & 2.2 \\
    \rowcolor{black!12}\textbf{Active subtotal} &  & \textbf{454} & \textbf{90.8} & \textbf{500} & \textbf{100.0} & \textbf{954} & \textbf{95.4} \\
    \midrule
    \rowcolor{black!17}\textbf{Overall} &  & \textbf{500} & \textbf{100.0} & \textbf{500} & \textbf{100.0} & \textbf{1,000} & \textbf{100.0} \\
    \bottomrule
  \end{tabularx}
  \vspace{7pt}\par
  \textbf{Panel B: Task-family mix and model outcomes}\par\vspace{3pt}
  \begin{tabularx}{\linewidth}{l >{\hsize=0.9\hsize\centering\arraybackslash}X >{\hsize=0.9\hsize\centering\arraybackslash}X >{\hsize=0.9\hsize\centering\arraybackslash}X >{\hsize=0.65\hsize\centering\arraybackslash}X >{\hsize=1.15\hsize\centering\arraybackslash}X >{\hsize=1\hsize\centering\arraybackslash}X >{\hsize=1.3\hsize\centering\arraybackslash}X >{\hsize=1.2\hsize\centering\arraybackslash}X}
    \toprule
    & \multicolumn{3}{c}{\textbf{Canonical task mix}} & \multicolumn{5}{c}{\textbf{Fixed comparison cohort: Pass@3 solved counts}} \\
    \cmidrule(lr){2-4}\cmidrule(lr){5-9}
    \textbf{Task family} & \textbf{Wave 1} & \textbf{Wave 2} & \textbf{Total} & \textbf{$n$} & \textbf{\shortstack{GPT-5.6\\Sol}} & \textbf{Grok 4.6} & \textbf{\shortstack{MAI-Code\\1.1 Flash}} & \textbf{\shortstack{Gemini\\3.7 Flash}} \\
    \midrule
    \rowcolor{black!8}Bug Fix & 179 & 176 & 355 & 180 & 27 & 24 & 9 & 1 \\
    Feature Addition & 220 & 260 & 480 & 236 & 26 & 24 & 3 & 0 \\
    \rowcolor{black!8}Test Generation & 39 & 4 & 43 & 24 & 16 & 4 & 0 & 0 \\
    API Migration & 29 & 17 & 46 & 25 & 5 & 5 & 1 & 0 \\
    \rowcolor{black!8}Security Repair & 33 & 43 & 76 & 35 & 5 & 5 & 0 & 0 \\
    \midrule
    \textbf{Overall} & \textbf{500} & \textbf{500} & \textbf{1,000} & \textbf{500} & \textbf{79} & \textbf{62} & \textbf{13} & \textbf{1} \\
    \bottomrule
  \end{tabularx}
  \caption{Industrial task composition and evaluation. Panel A shows 1,000 cases across anonymized repositories: R1--R2 are early repositories and R3--R11 the nine active repositories. Shares use the corresponding wave or full cohort as denominator. Panel B gives canonical family counts and Pass@3 solved counts on a separate fixed 500-case model comparison cohort.}
  \label{tab:app-product-repos}\label{tab:app-product-family-model}
\end{table}

\par\addvspace{6pt}\noindent
\begin{evidencecolumn}{.62}
The canonical cohort has two 500-case waves. GPT-5.6 Sol mean Pass@1 is 20.67\% and 1.40\%, and Pass@3 is 27.2\% and 2.0\% in Waves 1 and 2, respectively; overall values are 11.03\% and 14.6\%. Wave 1 contains 179 Bug Fix, 220 Feature Addition, 39 Test Generation, 29 API Migration, and 33 Security Repair tasks; Wave 2 contains 176, 260, 4, 17, and 43. Construction routes are 20/1/479 across Levels 1/2/3 in Wave 1 and 0/0/500 in Wave 2.

\end{evidencecolumn}\hfill%
\begin{evidencecolumn}{.35}
\centering

  \includegraphics[width=\linewidth]{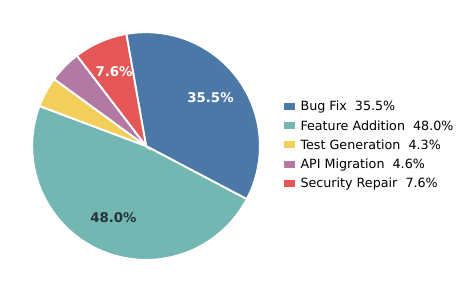}
  \sidecaption{figure}{Task-family composition of 1,000 industrial cases.}
  \label{fig:app-product-profile}
\end{evidencecolumn}
\par\addvspace{6pt}

Wave 2 selects 500 tasks from 16,601 discovered PR candidates: 2,736 meet static eligibility, 2,400 enter the construction reserve, 760 are attempted, and 703 pass strict Level 3 qualification. Source evidence includes 151 Azure DevOps PR exports and 349 verified Git commit records. The target check follows pass--fail--pass over healthy, challenge, and restored states; protected checks pass throughout. Three independent runs use a 12-turn and 1,800-second limit.

\subsection{Ablation: Cross-Model Comparison}

The fixed 500-case comparison set samples 250 cases from each wave: 180 Bug Fix, 236 Feature Addition, 24 Test Generation, 25 API Migration, and 35 Security Repair. The four models share the agent scaffold, prompt, tools, and hidden checks. GPT-5.6 Sol, Grok 4.6, MAI-Code 1.1 Flash, and Gemini 3.7 Flash solve 79, 62, 13, and 1 tasks at least once, yielding Pass@3 of 15.8\%, 12.4\%, 2.6\%, and 0.2\%. Family-level outcomes appear in Figure~\ref{fig:product-model} and Table~\ref{tab:app-product-family-model}. \textbf{The fixed evaluation exposes a broad range of model performance under one protocol.}
\endgroup

\FloatBarrier
\section{Qualitative Case Studies}\label{app:additional-analysis}

\subsection{Feature Addition Across Repository Evolution}

\begin{casebox}{Feature Addition Case: Towncrier PR 603}
\textbf{Historical maintenance request.} Add section-aware fragment creation.
The command-line interface accepts a section name, validates it against the
project configuration, supports interactive section selection, and resolves
the selected fragment directory through a shared path abstraction.

\textbf{Historical evidence.} The source PR adds the
\emph{section} option, introduces the section validation path, and centralizes
fragment-directory resolution. Its implementation patch spans four files,
13 hunks, and 160 changed lines.

\textbf{Healthy modern base.} The selected descendant still implements the
section-aware behavior. Its command path has evolved and now includes later
functionality such as the \emph{sub-issue} option, updated parameter plumbing,
and a refactored path helper. The four mapped target checks pass before
construction.

\textbf{Task shown to the coding agent.} ``Restore section-aware fragment
creation. The command must accept an explicit section, support the default and
interactive selection paths, and reject unknown section names while
preserving the current command interface.''

\textbf{Constructed task state.} Agent Reconstruction removes the modern semantic slice
that implements section selection and validation. It preserves
\emph{sub-issue}, the evolved command structure, and unrelated fragment
creation behavior. The resulting task patch spans two implementation files,
11 hunks, and 121 changed lines. The paired restoration patch reintroduces the
section-aware behavior in the modern code.

\textbf{Executable evidence.} Four target checks cover explicit selection,
default selection, interactive selection, and invalid names. Sixteen
regression checks protect existing fragment creation and later command
behavior. Target checks follow pass, fail, pass across the healthy, task, and
restored states. All regression checks pass in every state.
\end{casebox}

\begin{samepage}
This case captures the central distinction between reconstruction and source
checkout. A historical rollback would also remove behavior introduced after
PR 603. Change2Task preserves that later behavior and materializes the missing
feature inside the repository's current implementation context.\par
\end{samepage}

\subsection{Bug Fix with a Modernized Implementation}

\begin{casebox}{Bug Fix Case: Django Migration State}
\textbf{Historical maintenance request.} Preserve the
\emph{ProjectState.real\_apps} invariant. The constructor may receive
\emph{None} or a set, and a non-empty value must already satisfy the set
contract. The historical fix prevents permissive conversion from concealing
an invalid caller state.

\textbf{Historical evidence.} The source implementation patch changes seven
lines in one file and one hunk. Its target check distinguishes a valid set
from an invalid collection passed to the migration state.

\textbf{Healthy modern base.} The invariant remains in the evolved
migration-state implementation. The surrounding migration code, imports, and
test harness have changed, while the mapped target check and 32 adjacent
regression checks pass on the selected base.

\textbf{Task shown to the coding agent.} ``Correct
\emph{ProjectState.real\_apps} handling so that the migration state preserves
the documented set invariant without changing valid migration behavior.''

\textbf{Constructed task state.} Agent Reconstruction restores the earlier
permissive conversion in the modern constructor. The six-line task patch
touches one file and one hunk. Its restoration patch reinstates the invariant
without changing the modern method signature or adjacent migration logic.
The source-to-modern profile score is .960.

\textbf{Executable evidence.} The target check follows pass, fail, pass across
the healthy, task, and restored states. All 32 protected migration checks pass
throughout the lifecycle.
\end{casebox}

The case demonstrates that Agent Reconstruction also handles compact
maintenance obligations. It retains the historical trigger and expected
behavior while keeping the modern edit surface aligned with the seven-line
source change.

\section{Full Prompt Templates}\label{app:prompts}\label{app:prompt-templates}

The following prompts contain the complete instruction-bearing text used by
the corresponding stage. Placeholders enclosed by angle brackets are filled
from the frozen task record. Repository files are exposed through an isolated
worktree, while task evidence and verifier feedback are inserted into the
named fields.

\begin{promptbox}{Agent Reconstruction System Prompt}{c2tNavy}{c2tBlueBg}
\textbf{Role and mission.} You are the construction agent for Change2Task.
Your mission is to reconstruct, in a healthy modern base, the maintenance
condition represented by a historical PR. The resulting repository must
express the same developer-facing obligation under the modern implementation.
Construct the task state while leaving the maintenance task unresolved.
Preserve the modern repository structure and avoid a wholesale restoration of
historical source code.

\textbf{Provided evidence.}
\begin{enumerate}
  \item \textbf{Source request:} \promptfield{<issue\_or\_pr\_description>}.
  \item \textbf{Source change:} \promptfield{<historical\_implementation\_patch>}.
  \item \textbf{Modern base:} an isolated worktree at
        \promptfield{<healthy\_modern\_commit>}.
  \item \textbf{Behavior mapping:} \promptfield{<modern\_host\_files>},
        \promptfield{<symbols>}, and \promptfield{<call\_chain>}.
  \item \textbf{Verification contract:} \promptfield{<target\_checks>} and
        \promptfield{<protected\_regression\_checks>}.
  \item \textbf{Source profile:} changed files, hunks, lines, symbols, target
        checks, and regression checks.
  \item \textbf{Prior feedback, when present:}
        \promptfield{<apply\_or\_verification\_feedback>}.
\end{enumerate}

\begin{samepage}
\textbf{Required procedure.}
\begin{enumerate}
  \item Inspect the mapped files and trace the modern call path that implements
        the source PR's behavior. Confirm that the behavior remains present.
  \item Identify the smallest coherent modern semantic slice whose
        transformation recreates the source maintenance condition.
  \item Edit the isolated worktree to construct that condition. Preserve
        modern APIs, signatures, imports, wrappers, caching, and behavior
        introduced after the source PR unless they implement the target
        obligation itself.
  \item Review the resulting diff for completeness across affected call sites.
        Keep the transformation comparable to the source change in files,
        hunks, changed lines, symbols, and verification surface.
  \item Run focused syntax or build checks that are available within the
        worktree. Leave the final candidate applied for external verification.
\end{enumerate}\par
\end{samepage}

\textbf{Hard constraints.}
\begin{enumerate}
  \item Modify implementation files permitted by
        \promptfield{<allowed\_source\_paths>}.
  \item Preserve tests, fixtures, continuous integration files, generated
        files, dependency manifests, lock files, and repository metadata.
  \item Preserve protected adjacent behavior and current public interfaces.
  \item Do not add test-specific branches, hard-coded expected values, broad
        exception handling, dead code, stubs, or unrelated deletions.
  \item Do not collapse a multi-site obligation into one local shortcut or
        expand the task through unrelated refactoring.
\end{enumerate}

\textbf{Final response.} Return a concise
\promptfield{CONSTRUCTION\_REPORT} containing the modified files, the behavioral
slice reconstructed, the expected target effect, the protected behavior, and
the checks executed. The authoritative output is the clean unified diff left
in the worktree.

\textbf{Feedback rounds.} If a previous candidate failed, use the supplied
failure category and evidence to revise the transformation. Application
errors identify the rejected hunk and modern location. Target failures
identify the unmet task condition. Regression failures identify protected
checks. Fidelity feedback reports files, hunks, lines, symbols, and ratios.
Scope feedback identifies forbidden artifacts. Remove the previous
transformation before producing the next candidate.
\end{promptbox}

\begin{promptbox}{Matched PR Mirror Baseline Prompt}{c2tTeal}{c2tTealBg}
\textbf{Role.} You are given a historical PR patch and the corresponding files
from a healthy modern base of the same repository. Generate one candidate
patch that reverses the PR's change in the modern code.

\textbf{Inputs.}
\promptfield{<source\_request>},
\promptfield{<historical\_patch>},
\promptfield{<aligned\_modern\_files>},
\promptfield{<allowed\_source\_paths>}, and
\promptfield{<target\_and\_regression\_checks>}.

\textbf{Task.} Read the historical diff together with the complete modern file
context. Rewrite each affected modern file so that the behavior introduced or
repaired by the source PR is reversed. Retain modern code that is unrelated to
the source change, including later parameters, helpers, wrappers, and adjacent
features.

\textbf{Constraints.}
\begin{enumerate}
  \item Produce a source-code patch that applies to the supplied modern
        revision.
  \item Modify permitted implementation paths and preserve tests, fixtures,
        dependencies, build files, and repository metadata.
  \item Preserve current interfaces and unrelated behavior.
  \item Do not introduce test-specific logic, hard-coded outputs, broad
        exception handling, or unrelated cleanup.
\end{enumerate}

\textbf{Return format.} Return one unified diff and no explanatory prose. This
is a one-shot construction prompt. No verifier feedback or prior candidate is
provided. The candidate is evaluated under the same final verifier and
execution budget as Change2Task.
\end{promptbox}

\begin{promptbox}{Coding-Agent Evaluation System Prompt}{c2tOrange}{c2tOrangeBg}
\textbf{Mission.} Resolve the supplied repository task. Work directly in the
visible checkout, inspect relevant implementation and tests, execute available
commands, and leave a minimal solution in the worktree.

\textbf{Visible inputs.}
\begin{enumerate}
  \item task statement: \promptfield{<task\_statement>},
  \item repository checkout: \promptfield{<branch\_workspace>},
  \item visible source code, tests, documentation, and repository tools, and
  \item task-family output contract: \promptfield{<output\_contract>}.
\end{enumerate}

\textbf{Execution protocol.}
\begin{enumerate}
  \item Reproduce or localize the requested behavior using repository evidence.
  \item Inspect the relevant call path before editing.
  \item Implement the smallest complete change that satisfies the task.
  \item Run focused checks, then broader repository checks when feasible.
  \item Review the final diff for scope, completeness, and accidental changes.
\end{enumerate}

\textbf{Task-family contracts.}
\begin{enumerate}
  \item \textbf{Bug Fix:} repair the described defect while preserving passing
        behavior.
  \item \textbf{Feature Addition:} implement the requested behavior and its
        integration with the existing interface.
  \item \textbf{Test Generation:} add tests that exercise the requested
        behavior. The accepted patch is restricted to test artifacts.
  \item \textbf{API Migration:} replace obsolete source API usage with the
        target API and preserve externally observable behavior.
  \item \textbf{Security Repair:} remove the vulnerable behavior while
        preserving valid functionality under the supplied security oracle.
\end{enumerate}

\textbf{Acceptance condition.} The final repository state must satisfy hidden
target checks, protected regression checks, and the task-specific scope
contract. Construction traces, task patches, restoration patches, and hidden
checks are unavailable during the run.

\textbf{Final response.} Summarize the implemented change and the checks run.
The authoritative answer is the final worktree diff.
\end{promptbox}

\begin{promptbox}{Semantic-Alignment Judge Prompt}{c2tPurple}{c2tPurpleBg}
\textbf{Role.} Act as an independent judge of whether a reconstructed modern
task preserves the maintenance intent of its historical source task. Base the
comparison on behavioral obligations. Ignore textual patch similarity and
implementation style.

\textbf{Historical record.}
\promptfield{<historical\_request>},
\promptfield{<historical\_trigger>},
\promptfield{<historical\_expected\_behavior>},
\promptfield{<historical\_invariant>},
\promptfield{<historical\_affected\_api>}, and
\promptfield{<historical\_checks>}.

\textbf{Modern record.}
\promptfield{<modern\_task\_statement>},
\promptfield{<modern\_trigger>},
\promptfield{<modern\_expected\_behavior>},
\promptfield{<modern\_invariant>},
\promptfield{<modern\_affected\_api>},
\promptfield{<source\_to\_modern\_mapping>},
\promptfield{<contrastive\_checks>}, and
\promptfield{<restoration\_behavior>}.

\textbf{Decision procedure.}
\begin{enumerate}
  \item Compare the user-visible trigger and precondition.
  \item Compare the expected observable behavior and violated invariant.
  \item Check whether API or symbol changes are justified by repository
        evolution.
  \item Use positive, negative, boundary, adjacent, and restoration checks to
        detect weakened or shifted obligations.
  \item Assign exactly one label:
        \emph{exact intent match},
        \emph{intent-preserving adaptation},
        \emph{related but weakened},
        \emph{related but shifted},
        \emph{non-equivalent}, or
        \emph{insufficient evidence}.
\end{enumerate}

\textbf{Return format.} Return one JavaScript Object Notation (JSON) object
with fields
\promptfield{label}, \promptfield{trigger\_match}, \promptfield{behavior\_match},
\promptfield{invariant\_match}, \promptfield{mapping\_valid},
\promptfield{contrastive\_evidence}, and \promptfield{brief\_justification}.
The construction route, construction model, construction trace, profile score,
candidate ranking, and downstream agent outcomes are not provided.
\end{promptbox}

\FloatBarrier
\section{Limitations}\label{app:limitations}

\sys requires a traceable source PR, identifiable maintenance behavior, executable checks, and a healthy descendant hosting that behavior. Recovery rates therefore apply to the construction-eligible pool. Sparse or unstable checks narrow the observable contract; performance, service, interface, and hardware changes may need additional project-specific oracles.

The evaluation spans five task families from public Python and Java corpora and four agent configurations. Other languages, maintenance obligations, and agent interfaces may change recovery and solving behavior. Environment savings depend on task density per modern base, compute prices, storage policy, and retention period; Section~\ref{app:rq4-details} states the matched accounting boundary.